# Tuning the Dirac Cone of Bilayer and Bulk Structure Graphene by Intercalating First Row Transition Metals using First Principles Calculations

Srimanta Pakhira[†,‡] and Jose L. Mendoza-Cortes[*,†,‡]

*Condensed Matter Theory, National High Magnetic Field Laboratory (NHMFL), Scientific Computing Department, Materials Science and Engineering, High Performance Materials Institute (HPMI), Florida State University, Tallahassee, Florida, 32310, USA.*

E-mail: mendoza@eng.famu.fsu.edu

Phone: +1-850-410-6298. Fax: +1-850-410-6150

**Abstract**

Modern nanoscience has focused on two-dimensional (2D) layer structure materials which have garnered tremendous attention due to their unique physical, chemical and electronic properties since the discovery of graphene in 2004. Recent advancement in graphene nanotechnology opens a new avenue of creating 2D bilayer graphene (BLG) intercalates. Using first-principles DFT techniques, we have designed 20 new materials *in-silico* by intercalating first row transition metals (TMs) with BLG, i.e. 10 layered structure and 10 bulk crystal structures of TM intercalated in BLG. We investigated the equilibrium structure and electronic properties of layered and bulk structure BLG intercalated with first row TMs (Sc-Zn). The present DFT calculations show that the $2p_z$ sub-shells of C atoms in graphene and the $3d_{yz}$ sub-shells of the TM atoms provide the electron density near the Fermi level controlling the material properties of the BLG-intercalated materials. This article highlights how the Dirac point moves in both the BLG and bulk-BLG given a different TM intercalated materials. The implications of controllable electronic structure and properties of intercalated BLG-TM for future device applications are discussed. This work opens up new avenues for the efficient production of two-dimensional and three-dimensional carbon-based intercalated materials with promising future applications in nanomaterial science.

[*]To whom correspondence should be addressed

[†]Condensed Matter Theory, National High Magnetic Field Laboratory (NHMFL), Scientific Computing Department, Materials Science and Engineering, High Performance Materials Institute (HPMI), Florida State University, Tallahassee, Florida, 32310, USA.

[‡]Department of Chemical & Biomedical Engineering, Florida A&M University - Florida State University, Joint College of Engineering, Tallahassee, Florida, 32310, USA.

## Introduction

In honeycomb lattices like graphene, the existence of the Dirac point results from the planar trigonal connectivity of the sites and its sublattice symmetry.[1] Graphene is a wonderful material with many superlatives to its name. Monolayer graphene (MLG) is an infinite one atom thick honeycomb membrane of carbons in which each atom connects to three surrounding other carbons by $sp^2$ hybridized bonds, and it is often treated theoretically as a free-standing two-dimensional (2D) sheet. It has attracted great attention in nanomaterial science and nanotechnology since its first experimental fabrication in 2004,[2] due to its exceptional electronic properties linked to the Dirac physics of its low energy quasi-particles.[3,4] The presence of a Dirac Cone in graphene, which represents linear energy dispersion at the Fermi level, gives graphene massless fermions leading to various quantum Hall effects, ultra high carrier mobility, and many other novel phenomena and properties. MLG has been found to have excellent electrical, electronic and 2D layer material properties. Hindrances to using graphene in many applications, such as electronics have been due to the lack of a band gap. In modern digital electronic and nanoelectronic devices, MLG has the well known zero band gap issue[5] which makes a high on-off ratio difficult; deeming it unsuitable for transistors, which are the foundation of all modern electronic and digital devices.

Aside from the peculiar massless Dirac fermions that emerge in MLG, stacking graphene sheets on top of the other may result in very different energy-momentum relations,[6] expanding the electronic versatility of this 2D carbon system. In the simplest architecture, two graphene layers can be arranged in an AA configuration;[4] the physical properties of such bilayer graphene (BLG) are correlated with the stacking order and relative twist angle,[7,8] with each type possessing a unique $\pi$-electron landscape.[9] The AA-stacked BLG unit cell is constructed in such a way to have one atom in one layer exactly above one atom of the second layer of graphene.[4] Similar to MLG; BLG is an atomically thin 2D layer material that has generated extensive research effort in modern nanoscience in this decade. Unlike MLG which has an unusual Dirac Cone band structure at low energies,[4,10] BLG has a parabolic band material with zero band gap. A recent experiment showed that BLG presents an interesting case in terms of its electronic properties at low energy[11] leading to an unconventional quantum Hall effect which arises from the chiral nature of the charge carriers in BLG.[1,10] Its band structure around the K and K' points displays hyperbolic bands touching at the neutrality point if trigonal warping terms are neglected.[12] More interestingly, the low-energy electronic structure and high density of states (DOSs) can potentially lead to be an intriguing material in that its electronic structure can be altered by intercalating metal atoms between two graphene layers. BLG can be both the Dirac and the Schrödinger systems, depending on its stacking structure.[4] When BLG is in the Bernal AB-stacked form, in which the upper graphene is shifted in-plane by $(\vec{a} + \vec{b})/3$ with respect to the lower one, it forms a 2D

Schrödinger system having a parabolic band dispersion (where $\vec{a}$ and $\vec{b}$ are the unit vectors of graphene sheet).[4,13] On the other hand, when it is in the AA-stacked form, in which the lattice of two layers is stacked without lateral shift, it forms a 2D Dirac system having a linear band dispersion.[14] This difference between the AB- and AA-stacked BLGs causes some difference in their physical and electronic properties. Specially, the Berry Phase is different between the two systems; that of Dirac system is $\pi$, while that of Schrödinger system is $2\pi$. A *"Berry Phase"* is a phase angle (i.e., running between 0 and $2\pi$) that describes the global phase evolution of a complex vector as it is carried around a path in its vector space.

To fabricate practical devices based on graphene and BLG, it is essential to modify and control electronic properties and parameters such as the band structures, DOSs, sign and concentration of carriers as well as the band gap at the Dirac point. Despite the unique intrinsic properties of MLG and BLG layer materials, various modification methods have been applied to these materials that yield even more exciting outcomes. In fact, several attempts have been made to achieve material engineering by (i) introducing an external electric field[15,16]; (ii) depositing atoms or molecules on a graphene sheet[17]; (iii) alloying and hybridization[18]; and (iv) intercalation[4,19–21] to mention a few. Each of these categories provide unique perspectives and advantages in studying both the fundamental science as well as applications in 2D materials. Among a variety of graphene-based materials, graphene intercalation materials are formed by insertion of molecular or atomic layers with various chemical species between graphite layers.[4,19–24] In graphite, however, intercalation of guest atoms and molecules into graphite layers is known to considerably modify the electronic structure, leading to unique physical properties and technological applications such as superconductivity and rechargeable batteries.[21,23,25] Intercalated graphene materials with a periodic arrangement of an alternating sequence of intercalant and graphene layers exhibit a variety of exotic electronic properties ranging from superconductivity to magnetism.[19,20,26] Recent advancement in graphene nanotechnology opens a new avenue of creating few-layer graphene intercalates. Combining layers of 2D materials in particular sequences to form multilayer structures provides an avenue for the manipulation of mechanical and electronic properties and for creating heterostructure devices.[3,4,22,23,27]

As stated earlier, among all the approaches, the intercalation method is one of the best methods to control the material properties of graphene, BLG and graphite.[4,19,24,28,29] Alkali-metal intercalated graphite and graphene have been intensively studied for decades, where alkali metal atoms are found to form ordered structures at the hollow sites of hexagonal carbon rings. To the best of our knowledge, the intercalation of graphene or graphite was studied mainly by using alkali atoms or ions (Li, Na and K) which has many important applications in modern science[19–21] such as Li-ion or Na-ion. battery[28,30] However, intercalation of transition metal (TM) atoms into 2D layered structure and 3D bulk structure materials (such as bilayer graphene or graphite) has not been studied systematically and extensively yet. The few reported studies have shown that they provide rich electrical, material and chemical properties that are distinctly different from those of pristine materials.[4] In particular, intercalation into bulk BLG has attracted special attention, since graphite intercalation compounds show various fascinating physical properties such as superconductivity and magnetism.[19,20,26] Bui et al.[31] carried out a theoretical investigation on 2D BLG-Cr nanostructure material and they found that the $2p_z$ orbitals of graphene layers are aligned antiferromagnetically with respect to the Cr layer, but they did not study the electronic and material properties nor the details of the crystal structure. Schwingenschlögl and co-workers computationally investigated the electronic properties and magnetic behavior of a 3D graphitic network in $ABA$ and $AAA$ stacking with intercalated transition metal atoms (Mn, Fe, Co, Ni, and Cu) and they proposed that these spin-polarized TM-intercalated materials can be utilized in spintronic and nanoelectronic applications.[32] Xu and co-workers[33] theoretically investigated various ordered structures of 3D TM-intercalated BLG (where TM atoms are Sc and Ti) with biaxial strain and they computed the electronic and magnetic properties of these TM-intercalated BLG materials. Their results indicated that the strong interaction between TM atoms and graphene comes mainly from the hybridization between $p$ orbitals from C and $d$ orbitals from TM. BLG with different TM/carbon hexagons ratios (where TM = Ti, Cr, Mn, Fe) and insertion patterns were computationally studied by Zhang et al.,[34] and they showed a way to control magnetic and electronic properties of BLG.

Very recently, we investigated how the electronic properties of 2D BLG can be tuned by intercalation of Vanadium (V), Niobium (Nb), Tantalum (Ta) atoms and interestingly we found that the 2D layer structure BLG-V showed a Dirac Cone in its band structure.[4] We have also discussed why the Dirac Cone appears in the intercalated BLG materials,[4] considering only V-B TM atoms (such as Vanadium, Niobium and Tantalum) in the periodic table. Thus, the electronic structure and properties change as a function of the number of graphene layers as well as by intercalation of foreign atoms, i.e. TM atoms. Even though experimental and theoretical efforts on the TM intercalated BLGs are still limited; deep understanding toward modulating electronic properties on these materials is still lacking, which could be controlled by metal species or insertion of TM atoms and intercalation patterns. Inspired by TM functionalized single layer graphene, intercalating TM atoms into BLG are expected to enhance the stability by the $d$-orbitals of the TM and $p$-orbitals from graphene, and provide an alternate way to tune and control the electronic and material properties of BLG. In the present article, we have focused on the intercalation effects of all the first row transition metals on both the BLG and bulk-BLG showing the band structure and density of states (DOSs) near the Fermi energy level ($E_F$) as well as principles that control the electronic properties.

In order to design these novel devices (both layer and bulk structures BLG) it is critical to understand the structural and electronic properties of these intercalation compounds, specially inter-layer coupling of the bilayer graphene and transition metals in 2D (BLG-TM) and 3D (bulk-BLG-TM) structures. In this work, we use a periodic dispersion-corrected unrestricted hybrid DFT i.e. DFT-D[35–38] to show how the structure and material properties of both the layer and bulk structures of BLG have been changed by intercalating first row transition metals (TMs). We studied the band structures and density of states (DOSs) of first row transition metals intercalated in both the BLG and bulk-BLG. The present DFT-D study found a Dirac Cone at the K-point in the BLG-V among all the BLG-TM 2D layer materials while the other BLG-TM materials conserve the Dirac Point in their band structures. This study also found the Dirac point is very close to the Fermi energy ($E_F$) level in Mn and Cu intercalated bulk-BLG. Thus, we have further discussed how the Dirac Point is controlled and moved due to the TM intercalation in both BLG and bulk-BLG materials. This study may allow bilayer graphene, new avenues to develop 2D Dirac materials and modern electronic devices following recent trends in science and technology.

In this paper, we perform first-principles exhaustive investigation of first row TM atoms intercalation in BLG and bulk-BLG. We computationally designed a total of twenty TM intercalated materials: (i) ten TM-intercalated-BLG 2D layer materials (in short BLG-TM) and (ii) ten TM intercalated bulk structure BLG materials (in short bulk-BLG-TM). We have systematically studied the material properties of the aforementioned intercalated BLG materials in this study including equilibrium structure, stability and electronic properties. We observed in our computation that the Dirac point is at the Fermi energy level ($E_F$) of the 2D layer structure BLG-V (where V has the electronic configuration [Ar]$3d^3 4s^2$) formed the Dirac Cone at the K-point. However, when the BLG was intercalated by Ti ([Ar]$3d^2 4s^2$), where Ti has less *d*-orbital electrons compared to V, the Dirac point at the K-point moved down from the $E_F$ about 2 eV (see Scheme 1a). Similarly, the Dirac point at K moved below the $E_F$ in the 2D layer BLG-Fe where Fe ([Ar]$3d^6 4s^2$) has more *d*-orbital electrons compared to V as shown on the right side in Scheme 1a. Thus, these calculations reveal that the Dirac Cone and Dirac point are controlled by the *d*-orbital electrons of the TMs which are interacting with the *p*-orbital electrons of graphene in both the BLG-TM and bulk-BLG-TM materials. The Dirac features still exist in the TM intercalated bulk structure BLG material (bulk-BLG-TM); see Scheme 1b. For the bulk structure, the present study found that the Dirac point is very close to the ($E_F$) in the BLG-Mn and BLG-Cr as shown in Scheme 1b. In the bulk structure intercalated BLG structure i.e. bulk-BLG-TM, the Dirac point moved up and down depending on the TM atoms in BLG intercalated materials. Thus, in this study we present the factors that controls the Dirac point in BLG and bulk analog of BLG. The unit cell of the BLG-TM and bulk-BLG-TM are depicted in Scheme 2.

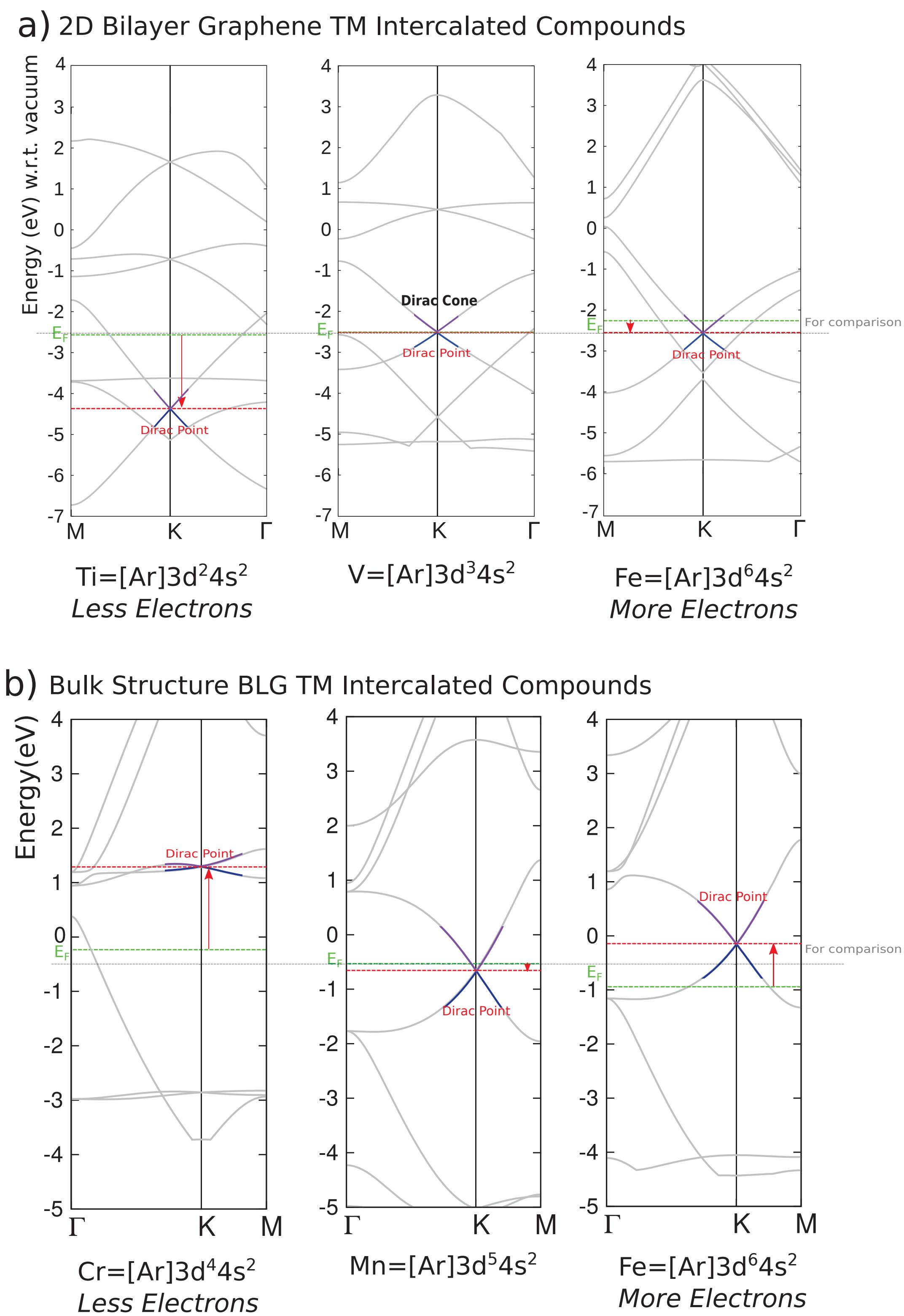

Scheme 1: Band structures of (a) BLG-V; BLG-Ti and BLG-Fe 2D-layered materials and (b) BLG-Mn, BLG-Cr and BLG-Fe 3D bulk structure materials. In these structures, the Dirac point is moved due to the different TM atoms i.e. changing the number of $d$ electrons in the system. In some cases, the Dirac point can touch the Fermi Energy level resulting into a Dirac Cone.

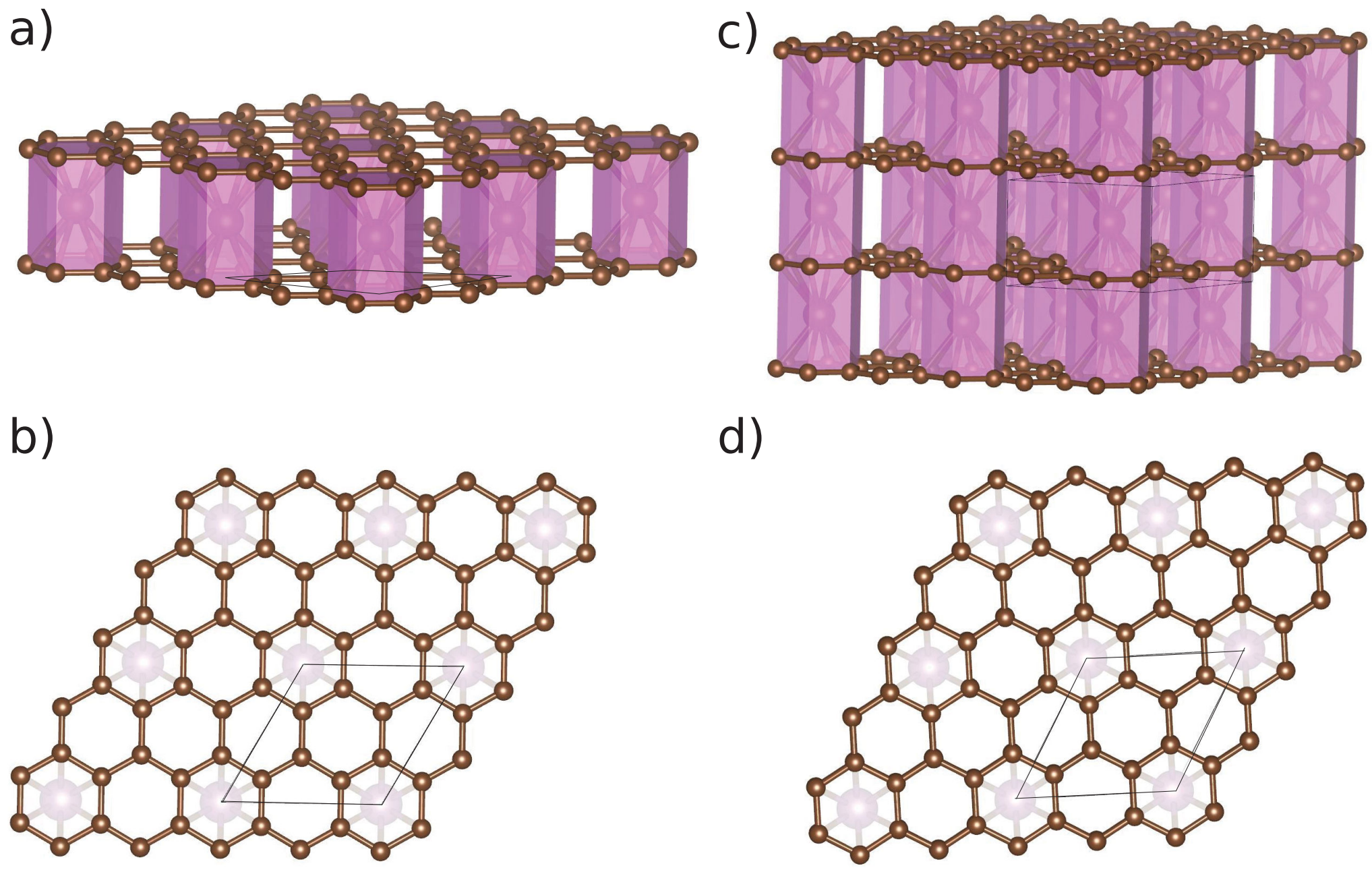

Scheme 2: Unit cells of the BLG-TM (a) side view, (b) top view; and bulk-BLG-TM (c) side view and (b) top view are shown here. The unit cells are highlighted by a parallelogram.

# Methods and Computational Details

First-principles calculations based on hybrid density functional theory (DFT) were used to perform all the periodic boundary computations as implemented in the *ab initio* CRYSTAL14 suite code, which makes use of localized Gaussian basis sets.[39] This approach differs from plane-wave codes (e.g. VASP, Quantum Espresso etc.) however both reaching similar results. However, for hybrid density functionals the localized Gaussian basis set codes are more naturally suited for solving the Hartree-Fock part of the solution. The equilibrium geometries of BLG-TM and bulk-BLG-TM (TMs: Sc, Ti, V, Cr, Mn, Fe, Co, Ni, Cu, and Zn) were obtained by the dispersion-corrected hybrid unrestricted DFT method, i.e. UB3LYP-D2, or DFT-D in short.[35–38,40,41] The semi-empirical Grimme's "-D2" dispersion corrections were added in the present calculations in order to incorporate van der Waals dispersion effects on the system.[38] This level of theory has shown to give correct electronic properties of 2D/3D materials.[4,42–44]. In the present computation, triple-zeta valence with polarization quality (TZVP) Gaussian basis sets were used for the C and all TM atoms.[45] 20 materials with BLG-TM as building blocks were designed including 10 BLG-TM and 10 bulk-BLG-TM materials: (a) BLG-Sc, (b) BLG-Ti, (c) BLG-V, (d) BLG-Cr, (e) BLG-Mn, (f) BLG-Fe, (g) BLG-Co, (h) BLG-Ni, (i) BLG-Cu and (j) BLG-Zn. Each one has been prepared by adding one TM atom in one unit cell between two graphene layers to construct the BLG-TM and bulk BLG-TM structures as shown in Figure 1 and Figure 2, respectively. The threshold used for evaluating the convergence of the energy, forces, and electron density was $10^{-7}$ a.u. for each parameter. The BLG-TM unit cell constructed this way contains one configuration, known as AA, where one atom is exactly above another atom of the other layer of graphene. In both the BLG and bulk-BLG, the TM atoms are inserted into the space between two graphene layers and form intercalated structures, which corresponds to $TM/C_8$ (graphene where the concentration of metal and C atoms is C:TM = 8:1 in brief $TM/C_8$) and results in the maximum capacity of bulk BLG during TM atoms intercalation. In both cases, one TM atom was intercalated per unit cell of BLG.

Integration inside of the first Brillouin zone was sampled on a 15 x 15 x 1 k-mesh grids for 2D layered materials and on 20 x 20 x 20 k-mesh grids for the bulk structures, with a resolution of around $2\pi$ x 1/60 $\text{Å}^{-1}$ for both the optimization and material properties calculations (band structures and DOSs). We have plotted the bands along a high symmetric k-direction, M − K − Γ − M for the bilayer materials and M − Γ − K − M − L − H − A for the bulk-BLG-TM materials in the first Brillouin zone. Electrostatic potential calculations have been included in the present computation for slabs/surfaces i.e. the energy is reported with respect to the vacuum for the bilayer materials. The band structures and DOSs of the BLG-TM and bulk-BLG-TM materials are shown in Figure 3 and Figure 4, respectively. The contributions of the sub-shells of the C and TM atoms (such as $p_x$, $p_y$, $p_z$, $d_{yz}$ etc.) in the total DOSs have been also computed for the BLG-TM bilayer and bulk-BLG-TM materials.

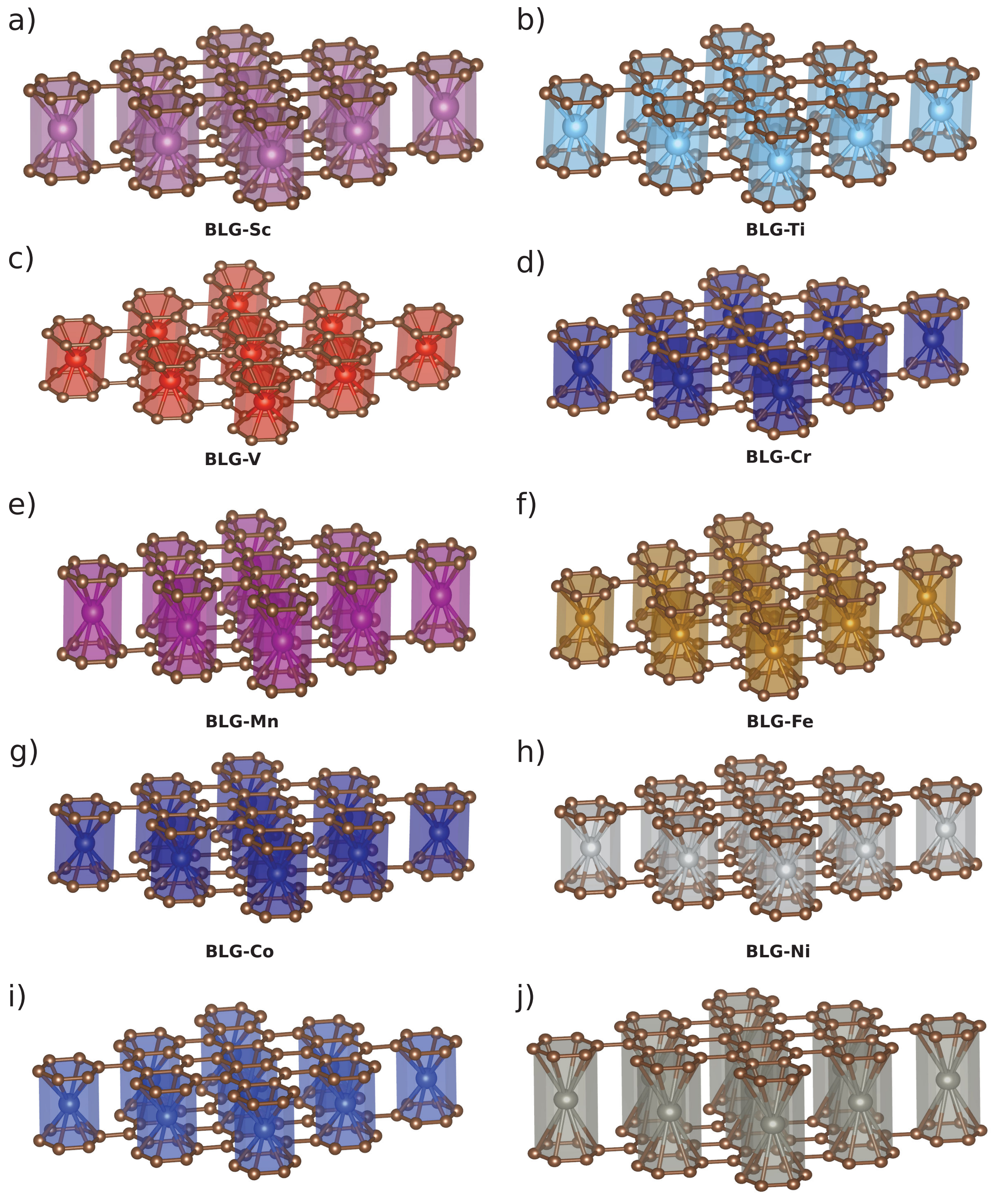

Figure 1: Equilibrium structures of BLG-TM; a) BLG-Sc, b) BLG-Ti, c) BLG-V, d) BLG-Cr, e) BLG-Mn, f) BLG-Fe, g) BLG-Co, h) BLG-Ni, i) BLG-Cu and j) BLG-Zn.

The conventional unit cells of the BLG-TM materials have a 2D characteristic in the $x$ and $y$ directions. The $z$ direction, on the other hand, is treated within a vacuum by employing a $\sim$ 500 Å length for the $z$ axis to accommodate the vacuum environment.

The mathematical expressions were used to calculate the binding energy of the pristine BLG and BLG-TM, $\Delta E_b$, are given below:

$$\Delta E_b = E_{BLG-TM} - E_{BLG} - E_{TM} \tag{1}$$

Where $E_{BLG-TM}$ is the energy of the BLG-TM, $E_{BLG}$ is the energy of the pristine BLG and $E_{TM}$ is the energy of TM atom.

## Results and Discussion

The optimized structures of BLG-TM and bulk-BLG-TM materials are shown in Figure 1 and 2, respectively, and the average C-TM and C-C bonding distances as well as the intercalation distances between two layers ($d$) are reported in Table 1. The electronic properties derived from the band diagram and the density of states are also shown in Table 1. The equilibrium lattice constants of the optimized structures of both the BLG-TM and bulk-BLG-TM materials are reported in the Supporting Information. The present DFT calculation found that the equilibrium average C-TM, C-C bond distances and intercalation distances ($d$) between two graphene layers were changed in both the 2D layer and bulk structure BLG-TM materials as depicted in Table 1. The average C-C bonds in graphene sheet in both the BLG-TM and bulk-BLG-TM materials have a resulting length $\sim$1.435 Å, which is longer that the C-C bond in an isolated monolayer graphene or bilayer graphene given our recent DFT calculations; 1.416 Å and 1.421 Å, respectively. The C-C and C-TM equilibrium bond distances and $d$ of the other BLG-TM layer materials are in agreement with the previous computational results.[31,33,34] The intercalation distances ($d$) in both the BLG-TM and bulk-BLG-TM are plotted in Figure 5 and the trend is similar for the bilayer and bulk analogs.

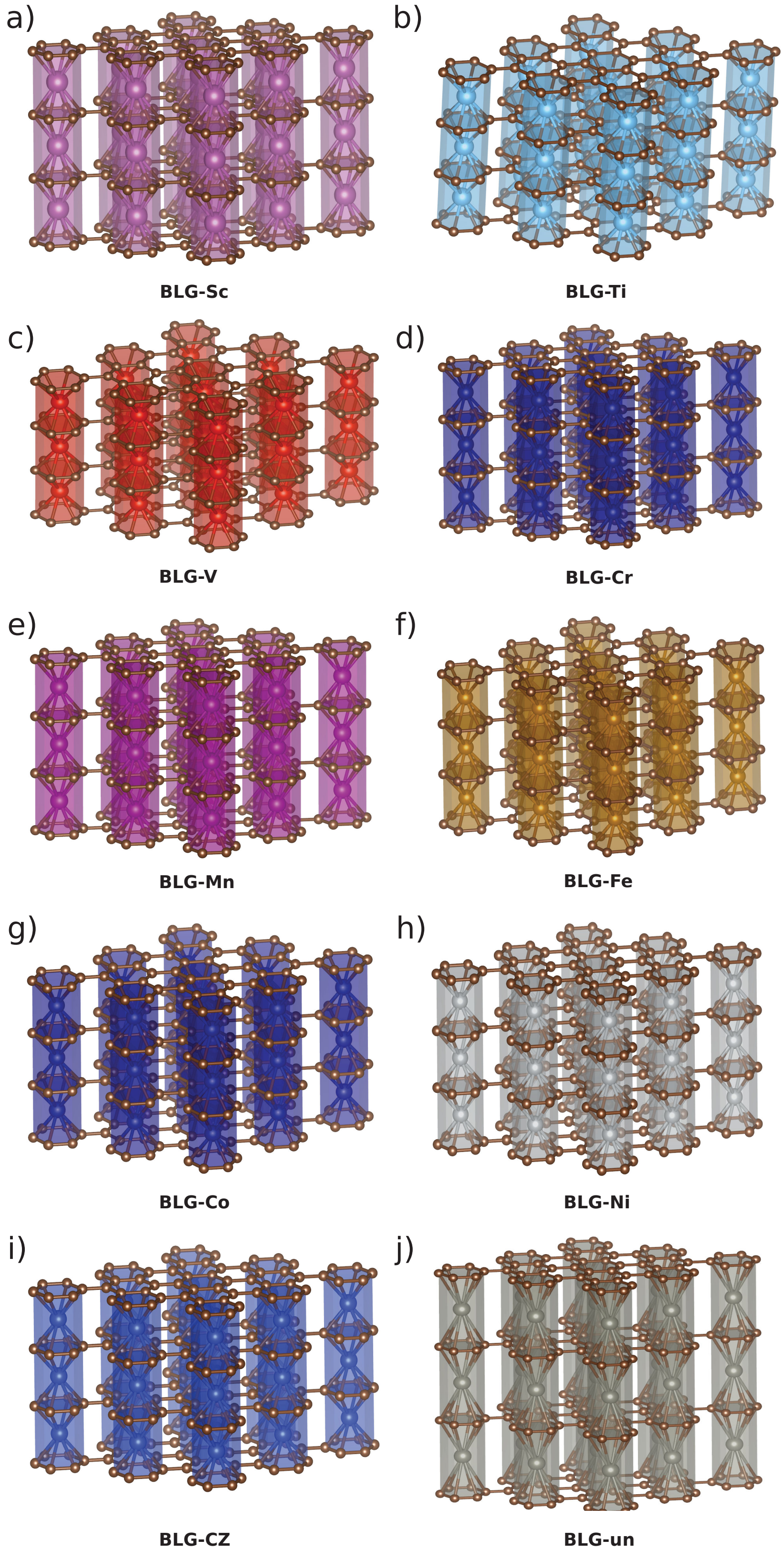

Figure 2: Equilibrium structure of bulk-BLG-TM intercalated materials; a) BLG-Sc, b) BLG-Ti, c) BLG-V, d) BLG-Cr, e) BLG-Mn, f) BLG-Fe, g) BLG-Co, h) BLG-Ni, i) BLG-Cu and j) BLG-Zn.

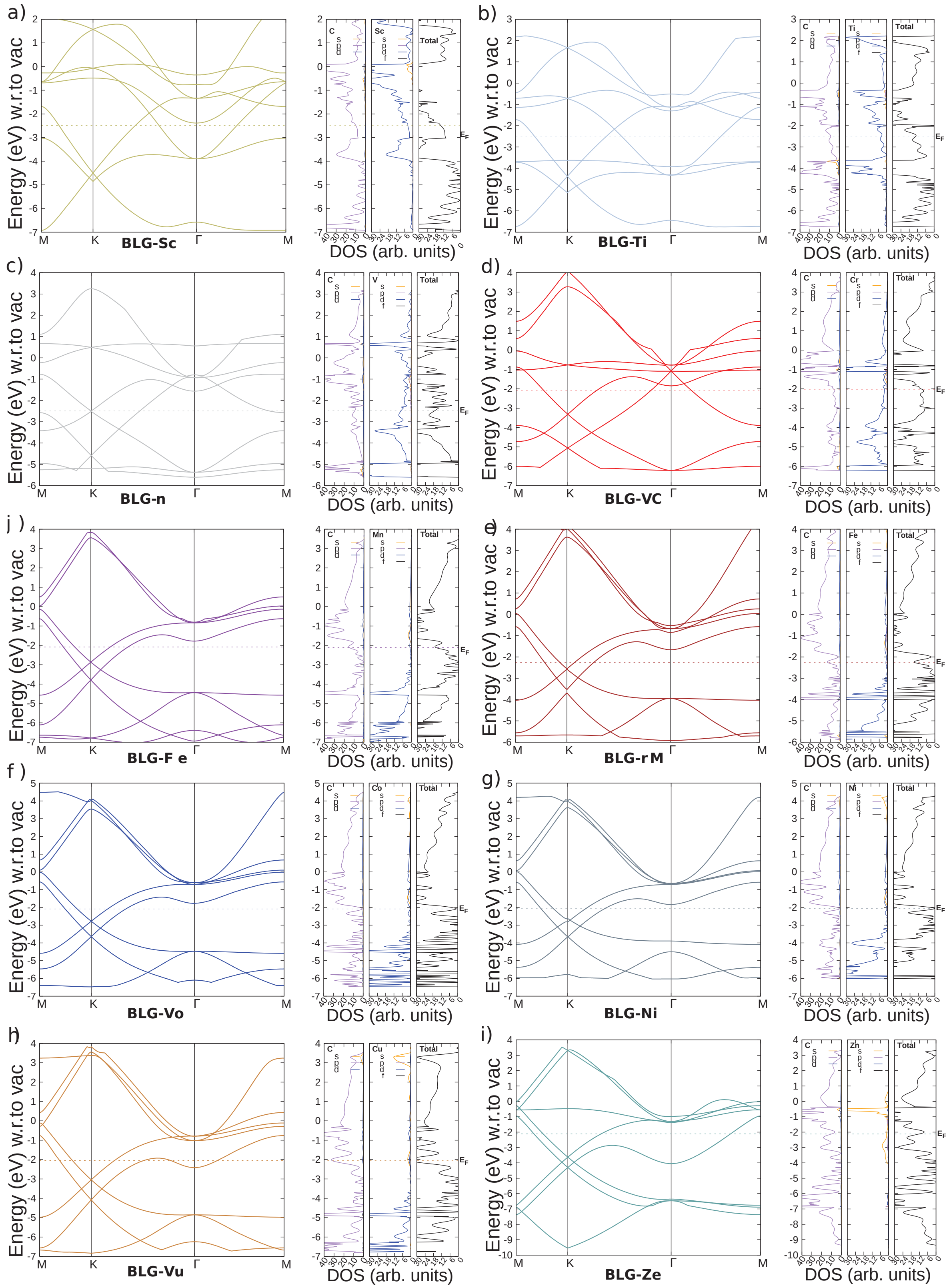

Figure 3: Band structure and DOSs of the BLG-TM materials; a) BLG-Sc, b) BLG-Ti, c) BLG-V, d) BLG-Cr, e) BLG-Mn, f) BLG-Fe, g) BLG-Co, h) BLG-Ni, i) BLG-Cu and j) BLG-Zn. The contribution of the individuals $s$ and $p$ orbitals of the C atoms, and $s$, $p$ and $d$ orbitals of the TM atoms were shown along with total DOSs.

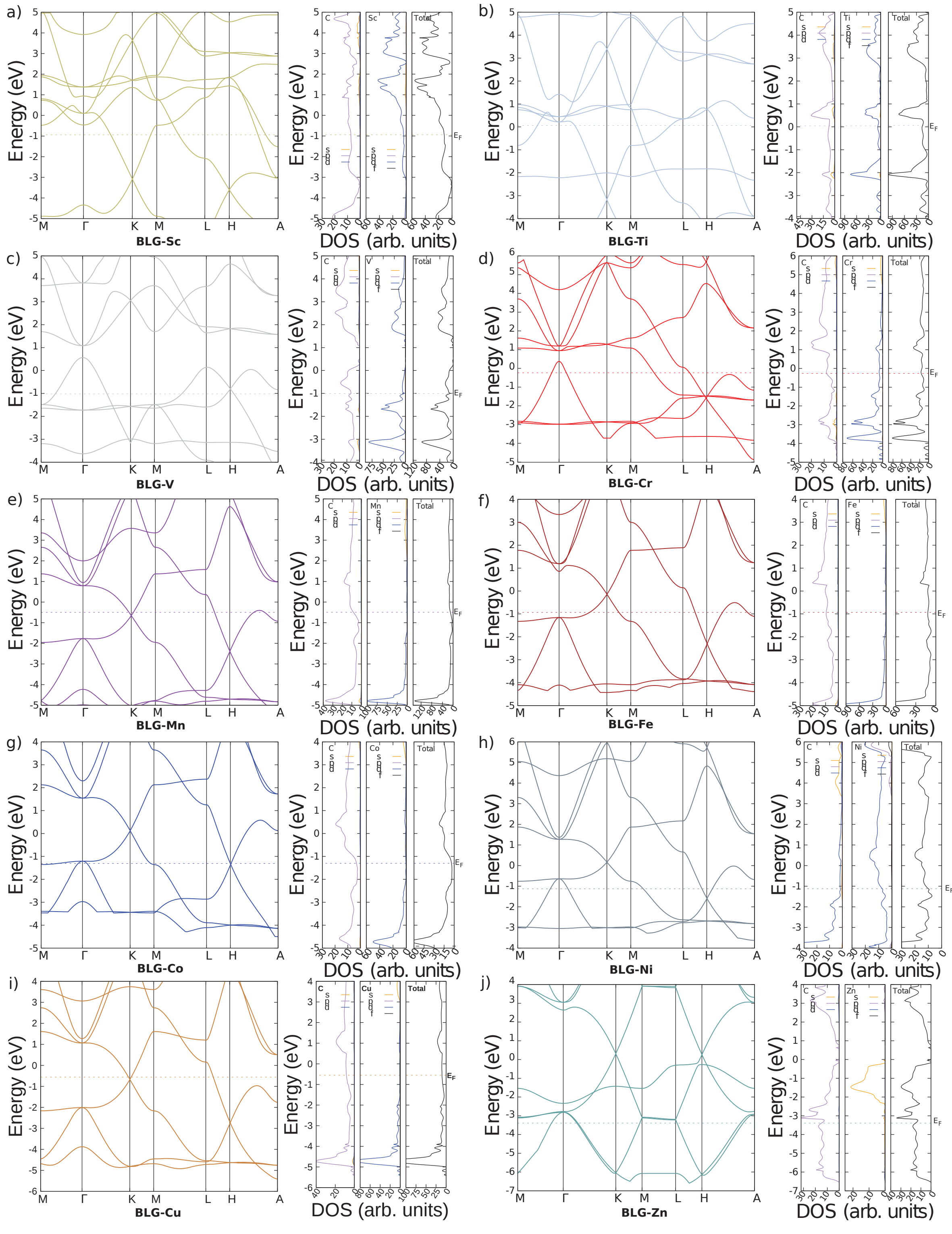

Figure 4: Band structure and DOSs of bulk-BLG-TM materials; a) BLG-Sc, b) BLG-Ti, c) BLG-V, d) BLG-Cr, e) BLG-Mn, f) BLG-Fe, g) BLG-Co, h) BLG-Ni, i) BLG-Cu and j) BLG-Zn. The contribution of the individuals $s$ and $p$ orbitals of the C atoms, and $s$, $p$ and $d$ orbitals of the TM atoms were shown along with total DOSs.

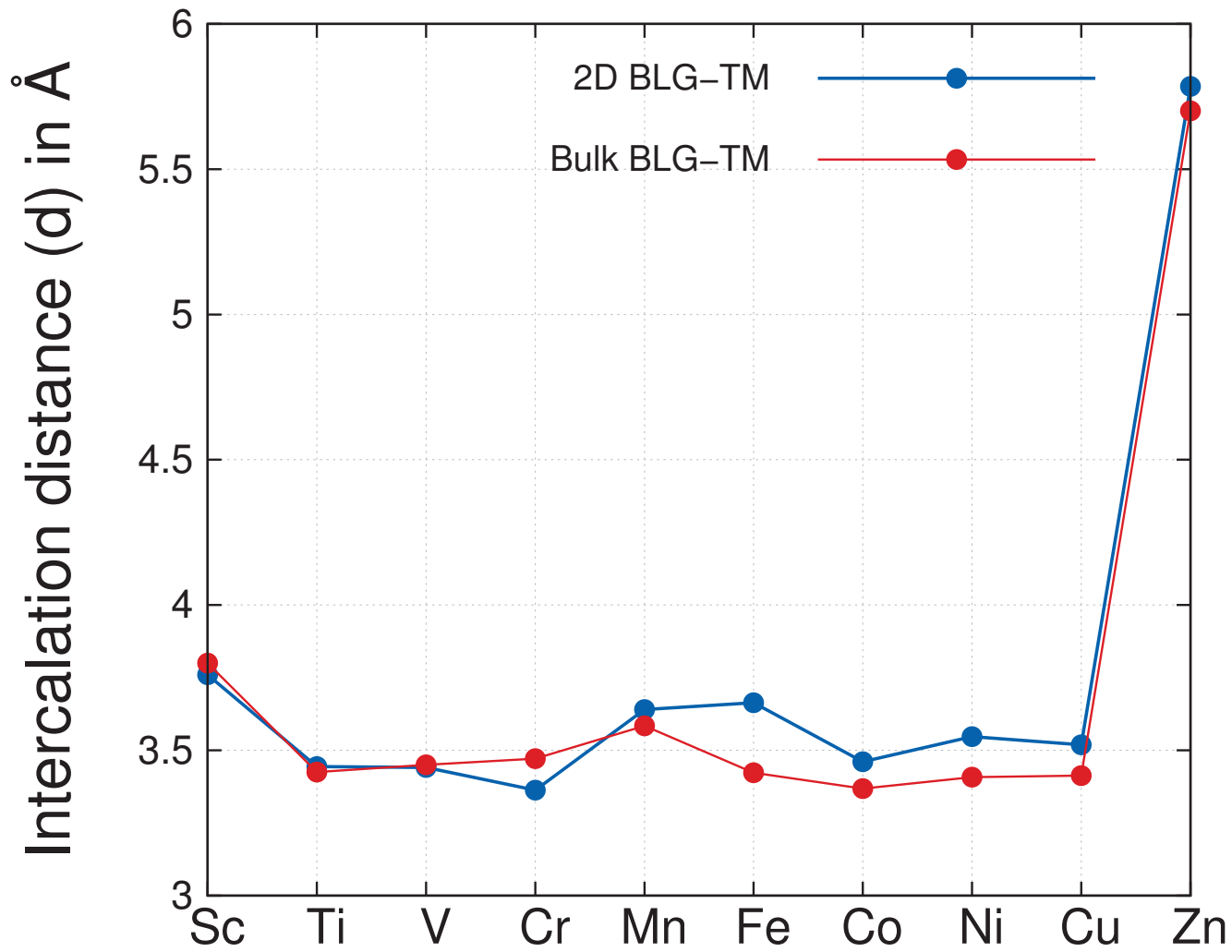

Figure 5: Equilibrium intercalation distance ($d$ in Å) for both the BLG-TM and bulk-BLG-TM materials.

To estimate the likehood of the synthesis of these materials, we have calculated the binding energy ($\Delta E_b$) of both systems, BLG-TM and bulk-BLG-TM. The binding energy was calculated using the mathematical expression in Eq. 1, and reported in Table 2. The present DFT study found that all the binding energies are negative indicating that the BLG-TM are more stable than BLG and TM alone. This calculation also reports that BLG-Zn is not a stable compound (both the layer and bulk structures) as $\Delta E_b$ is positive. Our computational study reveals that both the layer and bulk BLG-Ti materials have the most favorable binding energy among all materials studied here indicating it is the most stable compound, while the second most stable compound is BLG-Sc; see Table 2. The binding energy for both the BLG-TM and bulk-BLG-TM have been ploted and shown in Figure 6. This plot sows how the $\Delta E_b$ varies with the TM intercalated BLG materials and BLG-Ti is the most favorable material as it shows the lowest binding energy among all others studied here. Both the BLG-Mn and bulk-BLG-Mn have higher $\Delta E_b$ as depicted in Figure 6. This plot shows how the $\Delta E_b$ varies with the TM intercalated BLG materials, with the most stable compounds showing teh more negative energies.

Table 1: Equilibrium average C-TM and C-C bonding distances, intercalation distances ($d$) and electronic state for the BLG-TM and bulk-BLG-TM materials for the different first row transition metals (TM). The C-TM, C-C bond distances and $d$ were expressed in Å. The Mulliken spin population analysis on the TM atoms was also shown as 'Spin'.

| | BLG-TM | | | | | Bulk-BLG-TM | | | | |
|---|---|---|---|---|---|---|---|---|---|---|
| TM | C-TM | C-C | $d$ | Spin | State | C-TM | C-C | $d$ | Spin | State |
| Sc | 2.378 | 1.431 | 3.760 | 0.004 | Metal | 2.393 | 1.439 | 3.800 | 0.002 | Metal |
| Ti | 2.253 | 1.438 | 3.444 | 1.087 | Metal | 2.257 | 1.442 | 3.425 | 1.101 | Metal |
| V | 2.243 | 1.439 | 3.441 | 2.336 | Metal | 2.259 | 1.449 | 3.450 | 2.664 | Metal |
| Cr | 2.219 | 1.436 | 3.363 | 2.811 | Metal | 2.267 | 1.438 | 3.471 | 3.704 | Metal |
| Mn | 2.326 | 1.434 | 3.640 | 4.034 | Metal | 2.309 | 1.438 | 3.584 | 4.143 | Metal |
| Fe | 2.335 | 1.432 | 3.664 | 2.740 | Metal | 2.247 | 1.436 | 3.423 | 2.861 | Metal |
| Co | 2.252 | 1.431 | 3.461 | 1.671 | Semi-Metal | 2.221 | 1.441 | 3.368 | 1.574 | Metal |
| Ni | 2.294 | 1.430 | 3.547 | 0.619 | Semi-Metal | 2.233 | 1.430 | 3.408 | 0.304 | Metal |
| Cu | 2.274 | 1.429 | 3.519 | 0.002 | Metal | 2.238 | 1.432 | 3.413 | 0.000 | Metal |
| Zn | 3.222 | 1.416 | 5.785 | 0.000 | Metal | 3.182 | 1.413 | 5.701 | 0.000 | Metal |

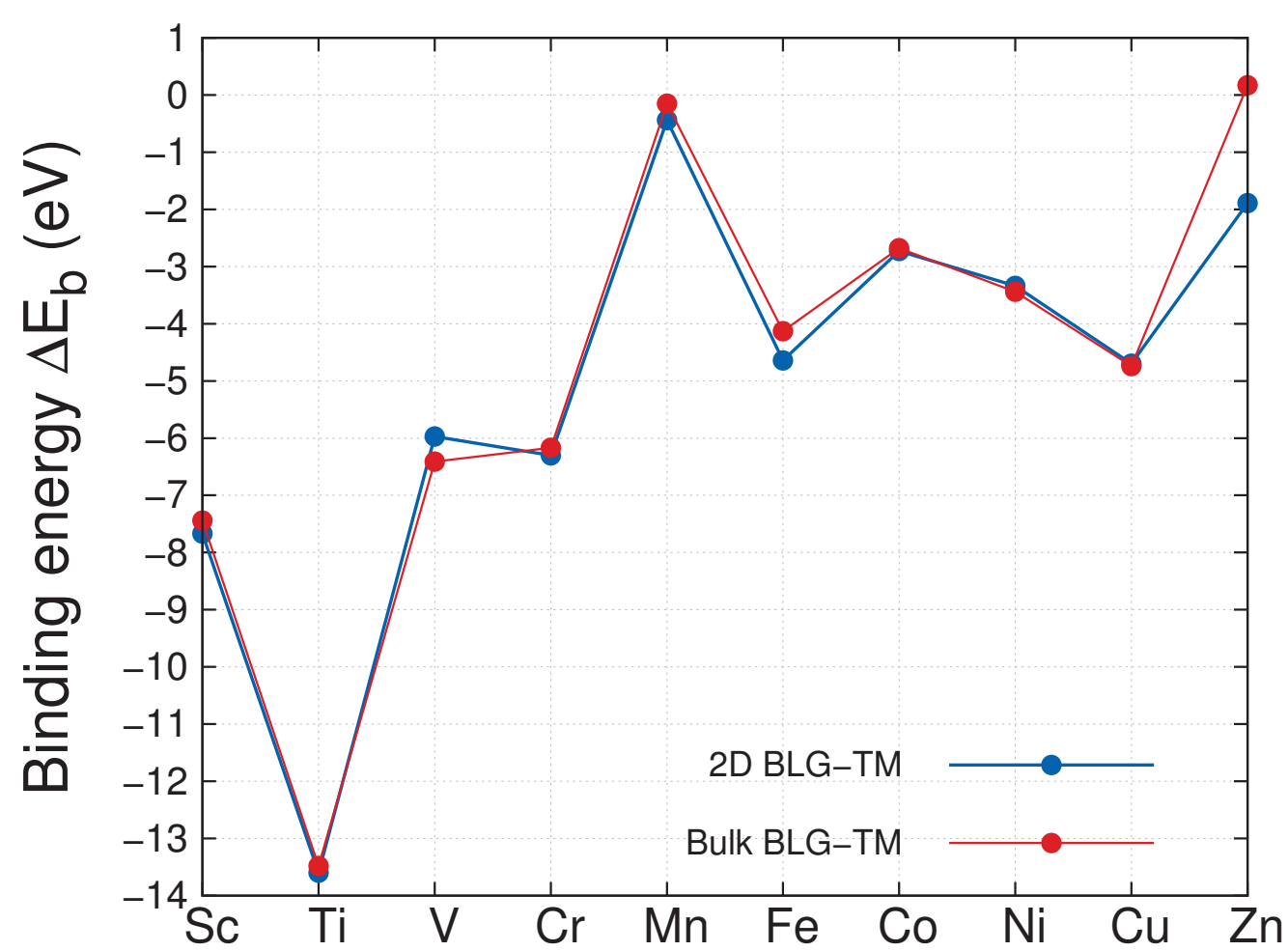

Figure 6: Binding energy ($\Delta E_b$ in eV) vs. both the BLG-TM and bulk-BLG-TM materials.

The pore surface area ($S_A$) of the BLG-TM and pore surface area ($S_A$), pore volume ($V_P$) and density ($\rho$) of the bulk-BLG-TM were estimated and reported in Table 2. Interestingly, our present DFT-D study found that BLG-TM have large surface area compared to bulk-BLG-TM, and plotted in Figure 7. This calculation indicated that the pore surface area of the BLG-TM was decreased while the larger TM atom was intercalated in BLG. It means that BLG-Sc has the largest pore surface area and BLG-Zn has the lowest pore surface area; see Table 2 and Figure 7. We also observed that the pore volume of the bulk-BLG-TM materials decreases as the size of the TM atoms increases.

Table 2: The binding energy $\Delta E_b$ of both BLG-TM and bulk-BLG-TM. Surface Area ($S_A$), Pore Volume ($V_p$), and Density ($\rho$) were estimated. The units of the $\Delta E_b$, $S_A$, $V_p$ and $\rho$ were expressed in eV, m$^2$ gm$^{-1}$, cm$^3$ gm$^{-1}$ and gm cm$^{-3}$. $S_A$ and $V_p$ were estimated from rolling a $H_2$ molecule with an initial solvent diameter of 2.80 Å, over the surface.

| | BLG-TM | | | | Bulk-BLG-TM | | | | | |
|---|---|---|---|---|---|---|---|---|---|---|
| TM | $\Delta E_b$ (eV) | $S_A$ (m$^2$ gm$^{-1}$) | $E_F$ (eV) | Dirac Point | $\Delta E_b$ (eV) | $S_A$ (m$^2$ gm$^{-1}$) | $V_P$ (cm$^3$ gm$^{-1}$) | $\rho$ (gm cm$^{-3}$) | $E_F$ (eV) | Dirac Point |
| Sc | -7.67 | 1101 | -2.487 | K | -7.44 | 620 | 0.198 | 5.05 | -0.987 | K, H |
| Ti | -13.60 | 1085 | -2.526 | K | -13.48 | 116 | 0.184 | 5.43 | 0.060 | K, H |
| V | -5.97 | 1069 | -2.490 | K | -6.41 | 124 | 0.182 | 5.48 | -1.007 | K, H |
| Cr | -6.30 | 1064 | -2.067 | K | -6.17 | 168 | 0.181 | 5.13 | -0.266 | K, H |
| Mn | -0.44 | 1048 | -2.084 | K | -0.15 | 358 | 0.183 | 5.46 | -0.493 | K, H |
| Fe | -4.64 | 1045 | -2.262 | K | -4.13 | 104 | 0.175 | 5.70 | -0.934 | K, H |
| Co | -2.73 | 1032 | -2.085 | K | -2.68 | 088 | 0.170 | 5.87 | -1.302 | K, H |
| Ni | -3.34 | 1031 | -2.047 | K | -3.44 | 097 | 0.172 | 5.81 | -1.118 | K, H |
| Cu | -4.70 | 1011 | -2.049 | K | -4.74 | 166 | 0.168 | 5.93 | -0.543 | K, H |
| Zn | -1.89 | 1007 | -2.113 | K | 0.17 | 1203 | 0.217 | 4.60 | -3.391 | K, H |

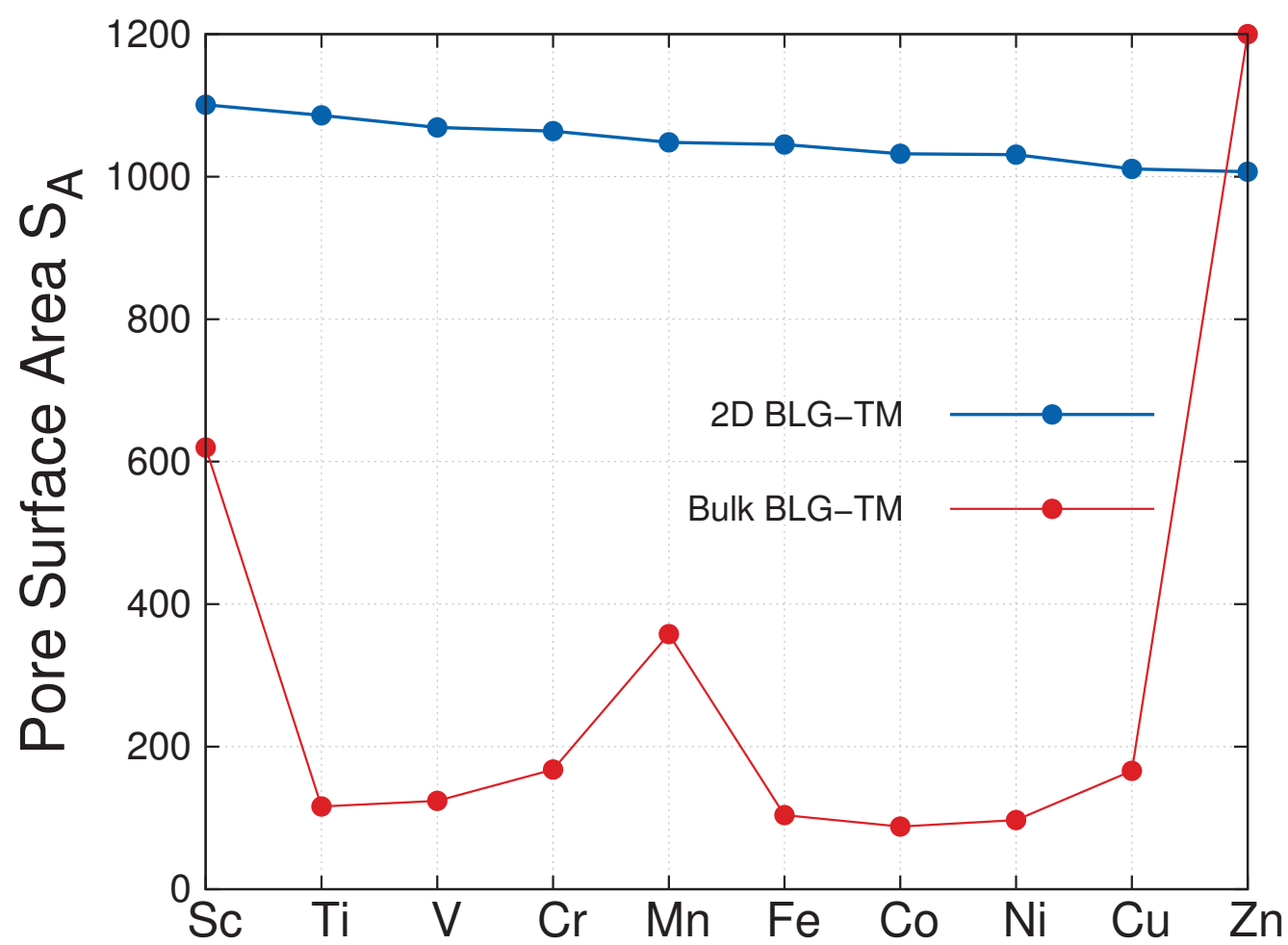

Figure 7: Pore surface area ($S_A$ in m$^2$ gm$^{-1}$) for both the BLG-TM and bulk-BLG-TM materials.

Obtaining the equilibrium structures of both the BLG-TM and bulk-BLG-TM materials,

we investigated how the electronic properties of the materials were changed by intercalating the first row transition metals. Very recently, we showed that the 2D AA-stacked BLG is a pure semiconductor with a small band gap $\sim 0.25$ eV and it has Dirac features. In the present study, we found that the material properties were drastically changed by intercalating first row TMs in the pristine AA-stacked BLG. For example, the band structure calculation showed that BLG-Sc and bulk-BLG-Sc behave as a metal (see Figure 3a and 4a) with large electron density around the Fermi Energy level ($E_F$). This is due to the contribution of the *p*-orbital electrons of the C atoms and the *d*-orbital electron of the Sc atom.

The electronic properties of the BLG-Ti and bulk-BLG-Ti structure materials were reported in Figure 3b and 4b, respectively. The present computation found that the intercalation distance $d$ and C-Ti bond distances were decreased whereas C-C bond distances increased in both the BLG-Ti and bulk-BLG-Ti materials compared to BLG-Sc (see Table 1) analog. Interestingly, we found a band crossing point i.e. Dirac point that still exists in both the BLG-Ti materials, but the valance band (VB) and conduction band (CB) are overlapped on each other making the materials a conductor. A large electron density was found around the $E_F$ due to VB and CB overlapping around the Fermi level as depicted in the "Total" DOSs. A detailed discussion of the contribution of individual sub-shells electron density of the C and TM atoms on the total density of states will be explained later in this article.

By intercalating a Vanadium (V) atom in BLG; the BLG-V and bulk-BLG-V materials were prepared. The present DFT-D calculations found that the intercalation distance $d$ almost remains the same in the bilayer structure while $d$ increasing by 0.025 Å in the bulk structure relative to BLG-Ti. The most interesting results were found BLG-V as depicted in Figure 3c as the addition of a single V atom intercalated in AA-stacked BLG (i.e. BLG-1V) yields a Dirac material. The present DFT calculation reveals that BLG-V has a Dirac Cone in its band structure, and other prominent graphene features. The DOSs around the Fermi energy level indicates metallic behavior. A detailed discussion about the reason of Dirac Cone in BLG-V material was reported very recently.[4] The electronic properties of the bulk-BLG-V were shown in Figure 4c. The Dirac Cone did not appear in the band structure of the bulk-BLG-V material, but the Dirac features and Dirac points remained at K- and H-points while valence and conduction bands are overlapped making it as a conductor as shown in Figure 4c. Interestingly, we found that the CB and VB touches each other just above the $E_F$ at the H-point (i.e. Dirac Point) making BLG-V a good conductor. A large electron density appears around the Fermi level which is coming mainly from the *p*-orbital electrons of C atoms in graphene and *d* orbital electrons of V, which provide the electron density around the $E_F$ as depicted in the right hand side of Figure 4c.

The structures and electronic properties were changed due to the addition of Cr and Mn atoms in bilayer and bulk-BLG. Both the bond distances (C-Cr and C-C) and intercalation distance *d* were decreased in the BLG-Cr in comparison with the BLG-V material. But for the case of bulk-BLG-Cr material, C-Cr and *d* distances were increased whereas C-C bond distances were decreased by 0.011 Å compared to bulk structure BLG-V. The C-Mn and *d* distances were increased in both the BLG-Mn and bulk-BLG-Mn materials compared to BLG-Cr and bulk-BLG-Cr; see Table 1. The Dirac Cone was moved down towards the VB after adding Cr and Mn in the BLG bilayer materials. In both the BLG-Cr and BLG-Mn materials, the VB and CB are overlapped below the Fermi energy level, making them conductors. Thus, the total DOSs calculations found a large electron density around the $E_F$ level, and the electron density is coming mainly from the *p*-orbitals of carbon of the graphene with a very small contribution from the *d*-orbital electrons as depicted in Figure 3d and Figure 3e, respectively. The electronic properties of the bulk-BLG-Cr and bulk-BLG-Mn materials are reported in Figure 4d-e. The valence and conduction bands of the BLG-Cr overlap and the Dirac point was pushed down towards the valence bands and it appeared around 1 eV below the $E_F$. The large electron density that has appeared around the $E_F$ suggests it is a conductor. The present DFT calculation found that another band crossing point, i.e. Dirac point, at the K-point which is very close to the $E_F$ in the bulk-BLG-Mn as shown in Figure 4e. Interestingly, it was observed that the *d*-orbital electrons of Mn are not providing any significant electron density in the total DOSs as shown in Figure 4e where as *p*-orbital electrons of the C atoms are significantly contributing the electron density on the total DOSs. Thus, the DOSs calculation indicated that graphene features are dominant in the bulk-BLG-Mn structure material.

The discussions about the geometrical structures and electronic properties of the BLG-TM and bulk-BLG-TM (Fe, Co, Ni, Cu and Zn) materials were reported in detail in the Supplementary Information. The band structures, DOSs and individual components of the *s*-, *p*-orbiatls electron density of the C atoms and the *s*-, *p*-, *d*-orbitals electron density of the TM (Fe, Co, Ni, Cu and Zn) atoms of the aforementioned materials are depicted in Figure 3f-j and Figure 4f-j, respectively. The most important findings about the material or electronic properties of the BLG-Fe, BLG-Co and bulk-BLG-Fe and bulk-BLG-Co materials were discussed here. The electronic properties of the Fe and Co atoms intercalated in BLG and bulk-BLG are shown in Figure 3f and in Figure 3g, respectively. Both the band structures and DOSs of BLG were changed due to the intercalation of the Fe and Co atoms in BLG, and the band structure calculations showed that the Dirac point is pushed down slightly from the $E_F$ by about 0.2 eV and 0.5 eV in the BLG-Fe and BLG-Co materials resulting in a conductor as depicted in Figure 3f and in Figure 3g, respectively. Interestingly, the present DFT computation found that the BLG-Co shows semi-metallic behavior like graphene, although the VB and CB are overlapped onto each other, but the total DOSs calculations show that the electron density is very close to zero at the $E_F$ (see Figure 3g) like monolayer graphene,[4] while BLG-Fe shows metallic behavior. The DOSs show that the total DOSs exactly follow the *p*-orbital electron density character of C atoms in graphene and *d*-orbitals electron has negligible contribution indicating that BLG-Co is a semi-metal. The DFT-D calculation found that VB and CB touch at $E_F$ yielding a Dirac Cone at the H-point in the bulk structure of the BLG-Co material, and at the same time VB and CB touch each on other at the K-point above around 1.3 eV above the $E_F$. This result means that the graphene features are dominant around the H-point in the bulk structure of BLG-Co. Thus we can predict that the bulk structure of the BLG-Co is a Dirac material whereas the 2D layer structure is a semi-metal with zero band gap.

The Dirac points at K-point and H-point pushed down in the BLG and bulk-BLG materials when they were intercalated by Ni and Cu atoms; see Figure 3h-i and 4h-i. Our present computation reveals that the total DOSs of the BLG-Ni is about zero at the $E_F$, whereas BLG-Cu has large electron density around the $E_F$. Therefore, it indicates that BLG-Ni has semi-metallic behavior as there is no electron density at or around the Fermi energy level. In the case of the layer structure BLG-Cu, a large electron density appears in the total DOSs around the Fermi level resulting a conductor due to the overlapping of VB and CB below the $E_F$ at the K-point as depicted in Figure 3i. The band calculations showed that the bulk-BLG-Cu the Dirac Cone appeared around the K-point; see 4i. Electron density, which follows the *p*-orbital character of C atoms of the graphene, occurs around the $E_F$ making it as a Dirac material. The Dirac point at K has pushed down below the $E_F$ level and the total DOSs showed large electron distribution around the $E_F$ of the BLG-Zn due to overlap of valence and conduction bands; see Figure 3j. Similarly, the VB and CB of the bulk-BLG-Zn overlap above the $E_F$ level resulting in a large electron density around the $E_F$. The present DFT calculation reveals that the total density of states of both the materials (bilayer and bulk structure) follow the *p*-orbital electron density, which also causes the large electron density around the Fermi level making them as a conductor.

We have further investigated in detail the contribution of the individual sub-shells electrons of the *p*-orbitals of C atoms and *d*-orbitals of TM atoms in the total DOSs of all the bilayer and bulk structure BLG studied here. The individual components of the sub-shells electron density (i.e. $p_x$, $p_y$, $p_z$, $d_{z^2}$, $d_{xy}$, $d_{yz}$, $d_{xz}$, and $d_{x^2-y^2}$) of all the bilayer and bulk structure BLG are shown in Figure 8 and Figure 9, respectively. The present study reveals that the $p_z$ sub-shell of carbon is the main contributing components, which provides the large electron density in the total DOSs of all the bilayer and bulk structure BLG materials. This calculation shows that the total DOSs follow the same trace of the $p_z$ sub-shell of carbon atoms in graphene. Among all the $d$-sub-shells of TMs, in most of the cases, mainly $d_{xy}$ and $d_{yz}$ provide the electron density around the $E_F$ level from the transition metals side in the total DOSs as shown in Figure 8 and Figure 9, respectively. Of course there are some exception, for example: bulk structure of BLG-Sc showed that $d_{z^2}$ is another contributing component in total DOSs along with $d_{xy}$ and $d_{yz}$ as shown in Figure 9a.

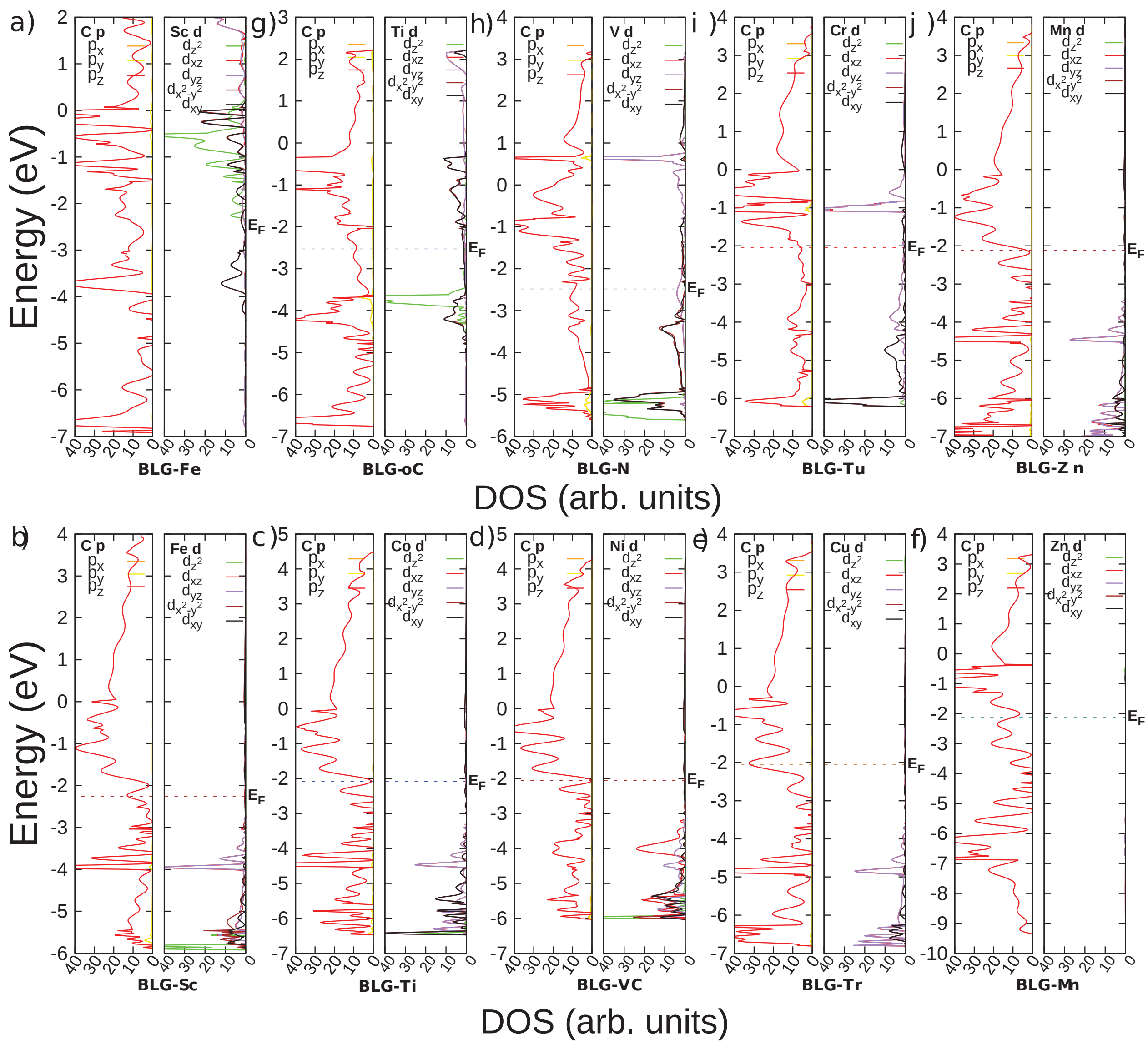

Figure 8: Sub-shells DOSs (i.e. the density of states of the $p_x$, $p_y$ and $p_z$ of the $p$-orbital of C atoms, and the $d_{z^2}$, $d_{xz}$, $d_{yz}$, $d_{x^2-y^2}$, and $d_{xy}$ of the $d$-orbital of TM atoms) of the BLG-TM materials; a) BLG-Sc, b) BLG-Ti, c) BLG-V, d) BLG-Cr, e) BLG-Mn, f) BLG-Fe, g) BLG-Co, h) BLG-Ni, i) BLG-Cu and j) BLG-Zn.

The present DFT calculations reveal that the TM atoms are making 2D layer and bulk structure BLG conductors or metals except for BLG-Co and BLG-Ni both the 2D layer and the bulk structure materials, which show semi-metallic behavior. The Dirac point and Dirac Cone move depending on the type of the transition metals intercalation for both the intercalated BLG-TM and bulk-BLG-TM materials. This study found that the Dirac point pushed down below the $E_F$ level in the 2D intercalated BLG except BLG-V, which showed the Dirac Cone in its band structures. The reasons for the Dirac Cone were discussed in detail in our previous study.[4] Different TM atoms intercalated BLG-TM 2D layer and bulk structure materials showed different electronic properties, which is completely dependent on the transition metal atoms. One of the important properties of the graphene, BLG and graphite is a semi-metal transition and their electronic properties can be controlled by TM intercalation, which we discussed in our present study. Very recently, Banerjee and his co-workers[23] experimentally showed that graphene intercalated compounds (i.e. doped multilayer graphene nanoribbons) paves the way for graphene as the next generation interconnect materials for a variety of semiconductor technologies and applications. They also computed the band structure calculations of the $FeCl_3$ intercalated ML-GNR material where the Dirac point moved above the Fermi energy level around 0.68 eV, which is very useful in interconnect technologies in integrated circuit (IC).[23] In general, Cu-based interconnects employed in a wide range of integrated circuit products are fast approaching a dead-end due to their increasing resistivity and diminishing current carrying capacity with scaling, which severely degrades both performance and reliability. In the present study, we found that the 2D layer structure BLG-V has a Dirac Cone, thus it might be useful for the next-generation IC and interconnect technology. The interesting thing is that the Dirac point appears in Cr, Mn, Fe, Co, Ni, Cu and Zn intercalated bulk structure BLG at the K-point, which have a similar type of band structure to the $FeCl_3$ intercalated ML-GNR material, and the Dirac point is very close to the Fermi level of the Cu and Mn intercalated bulk structure BLG. Thus these results indicate that these transition metal intercalated bulk structure BLG materials might be also useful for the next-generation IC and interconnect technology and they can potentially remove the Cu-based interconnect technology if these kind of materials can be prepared experimentally.

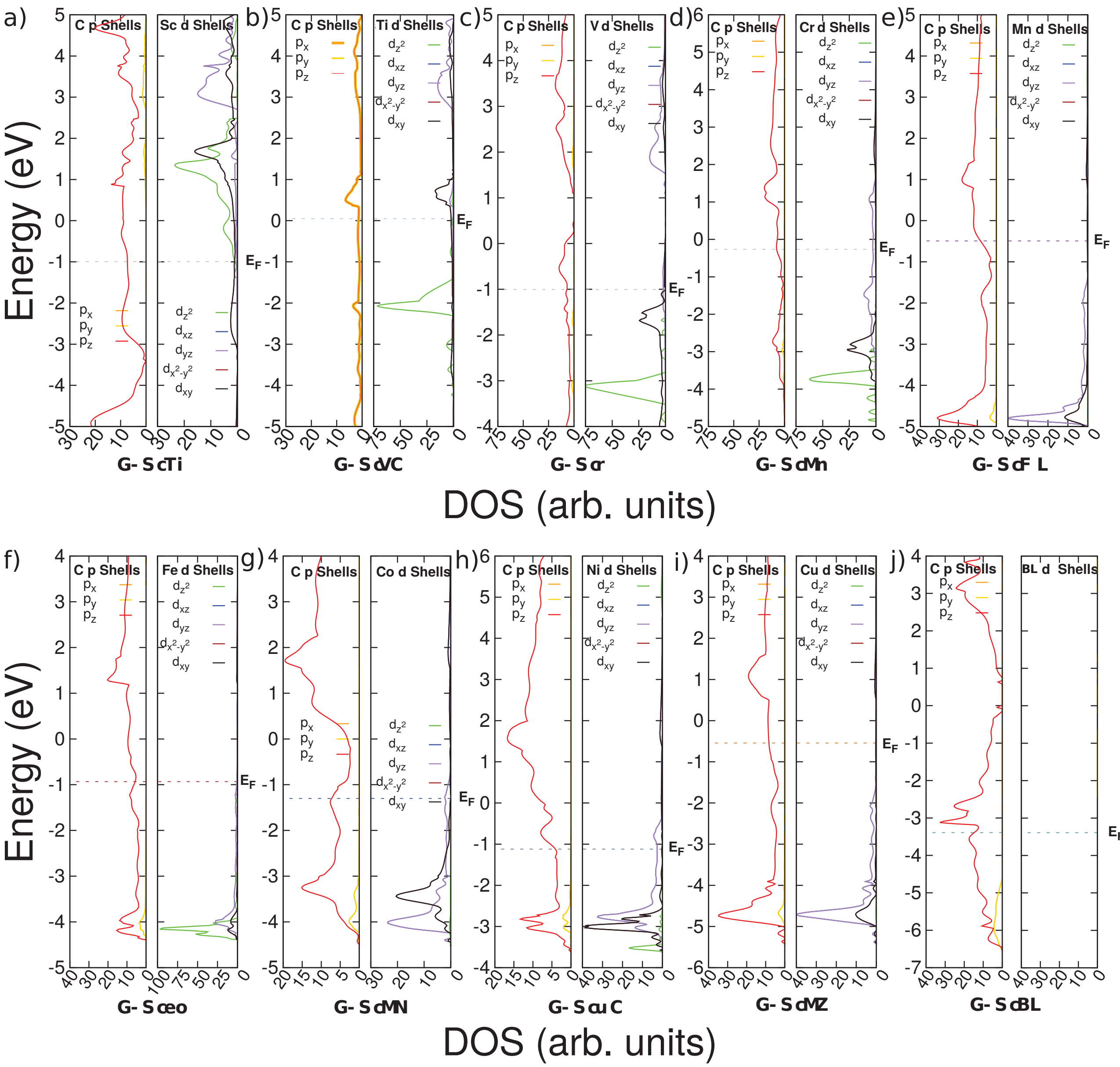

Figure 9: Sub-shells DOSs (i.e. the density of states of the $p_x$, $p_y$ and $p_z$ of the $p$-orbital of the C atoms, and the $d_{z^2}$, $d_{xz}$, $d_{yz}$, $d_{x^2-y^2}$, and $d_{xy}$ of the $d$-orbital of the TM atoms) of bulk-BLG-TM materials; ; a) BLG-Sc, b) BLG-Ti, c) BLG-V, d) BLG-Cr, e) BLG-Mn, f) BLG-Fe, g) BLG-Co, h) BLG-Ni, i) BLG-Cu and j) BLG-Zn.

# CONCLUSION

In summary, first row transition metals (Sc to Zn) intercalated 2D layer as well as bulk structure bilayer graphene have been investigated using first-principles hybrid density functional theory calculations. The equilibrium geometry of both the layer and crystal nanostructure TM intercalated materials were considered here and their electronic properties (band structures and DOSs) were computed by the DFT-D method. The individual components of the sub-shells of the $p$-orbitals (i.e. $p_x$, $p_y$ and $p_z$) of C atoms and the $d$-orbitals (i.e. $d_{z^2}$, $d_{xz}$, $d_{yz}$, $d_{x^2-y^2}$, and $d_{xy}$) of the TM atoms, which are taking part in the total electron density in DOSs have also reported along with the DOSs calculations. Among all the 2D layer nanostructure materials, the present study found only the BLG-V material has a Dirac Cone in its band structure. Two-dimensional first row transition metal atom intercalation in bilayer graphene opens new possibilities to engineer two-dimensional properties of intercalates as well as transition metal atom intercalation in bulk structure BLG also opens up the possibility of a general understanding of the intercalation behavior of BLG and the next generation interconnect material for a variety of carbon-based semiconductor technologies, integrated circuits and applications. Thus we showed a new way, transition metal intercalation which is a promising approach to control the electronic properties of BLG. This study will enhance the recent advancement in graphene nanotechnology, semiconductor technology, interconnect technology and opens a new avenue of creating both 2D layer as well as bulk structure bilayer graphene intercalates. Our results show that the interaction of TM atoms and bilayer graphene is so strong and that interesting electronic properties are identified.

We can say that the transition metal intercalation in both 2D layer and bulk structure bilayer graphene could also offer an interesting research approach for the creation of 2D layer structure and bulk crystal structure materials which have a potential application in nanomaterial science and nanotechnology. This study provides good insights into how to construct the 2D layer and crystal structure materials by TM intercalation between two layers graphene as a model system for computational analysis and how to control the material and electronic properties of BLG. Our results will certainly provide a new insight into the field of graphene intercalation science and nanomaterial future applications.

## Acknowledgement

This work was supported by start up funds from Florida State University (FSU). S.P. and J.L.M-C. gratefully acknowledges the support from the Energy and Materials Initiative at FSU. The authors thank the High Performance Computer cluster at the Research Computing Center in FSU, for providing computational resources and support. S.P. is grateful to Kevin P. Lucht from FSU for helpful discussions and guidance with computational resources.

## Supporting Information Available

The supporting Information is available free of charge on the ACS Publications website.

Section 1: Detail Discussions of the Structure and Electronic Properties of the BLG-TM and bulk-BLG-TM (where TM: Fe-Zn).

Section 2: Optimized Structures (.cif format) of the 2D layer structure BLG-TM materials.

Section 3: Optimized Structures (.cif format) of the 3D bulk crystal structure BLG-TM materials (i.e. TM intercalated BLG). This material is available free of charge via the Internet at `http://pubs.acs.org/`.

# Graphical TOC Entry

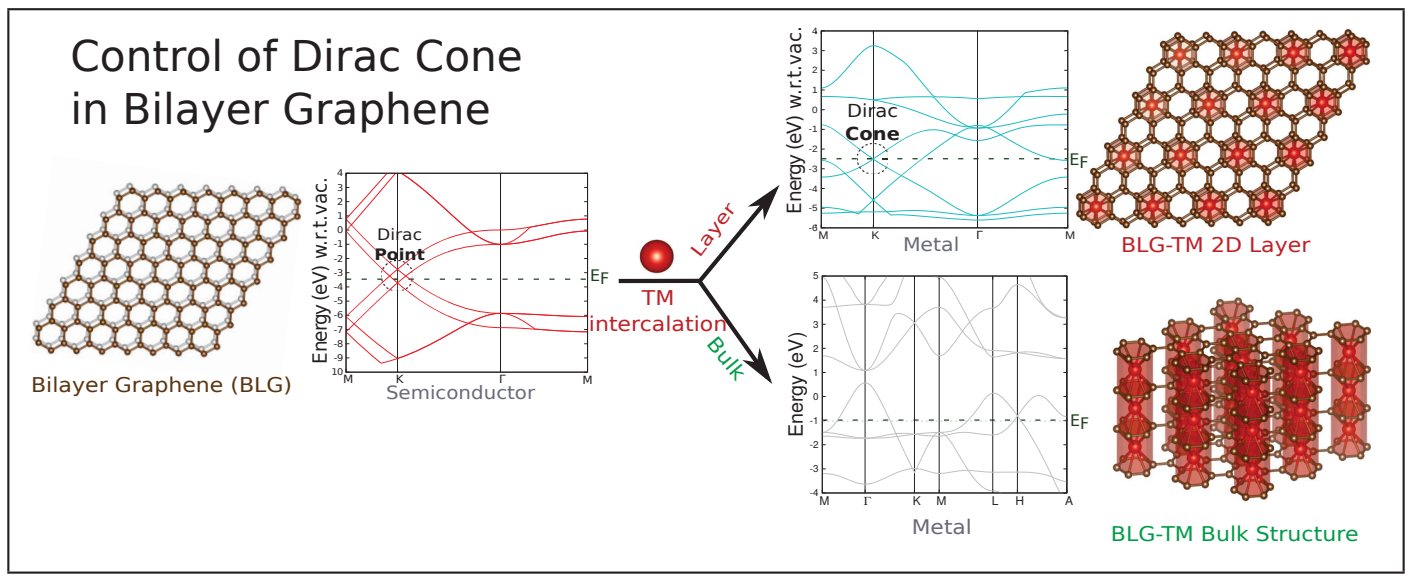

# Supplementary Information

## Tuning Dirac Cone of Bilayer and Bulk Structure Graphene by Intercalating First Row Transition Metals using First Principles Calculations

Srimanta Pakhira[1,2], Jose L. Mendoza-Cortes*[1,2]

[1] Condensed Matter Theory, National High Magnetic Field Laboratory, Scientific Computing Department, Materials Science and Engineering, Florida State University (FSU), Tallahassee, Florida, 32310, USA

[2] Department of Chemical & Biomedical Engineering, FAMU-FSU Joint College of Engineering, and High Performance Materials Institute (HPMI), Florida State University, Tallahassee, Florida, 32310, USA.

E-mail: mendoza@eng.famu.fsu.edu

# Contents

# 1 Detail Discussions of the Structure and Electronic Properties of the BLG-TM and bulk-BLG-TM (where TM: Fe-Zn)

The geometrical structures as well as electronic properties of the BLG and bulk-BLG materials were changed due to TM (TM: Fe, Co, Ni, Cu and Zn) intercalation. The equilibrium C-Fe bond distance and intercalation distance $d$ of the BLG-Fe were increased by an amount 0.009 Å and 0.024 Å respectively compared to the BLG-Mn. Similarly, the equilibrium bond and intercalation ($d$) distances were decreased in the bulk-BLG-Fe; see Table 1. The equilibrium C-Co, C-C and $d$ were also changed in both the BLG-Co and bulk-BLG-Co materials. But in the case of both the BLG-Co and bulk-BLG-Co, the bond distances and intercalation distance ($d$) were decreased in comparison with BLG-Fe and bulk-BLG-Fe. The electronic properties of Fe and Co atoms intercalated 2D layer structure BLG and bulk structure materials are shown in Figure 3f and in Figure 3g, respectively. The band structures and DOSs of BLG were changed due to the intercalation of Fe and Co atoms as the band structure calculations showed that the Dirac point is pushed down slightly from the $E_F$ level by about 0.2 eV and 0.5 eV in the BLG-Fe and BLG-Co materials resulting in a conductor as depicted in Figure 3f and in Figure 3g, respectively. A large electron density appears around the $E_F$ level in the total DOSs of the BLG-Fe due to the overlapping of VB and CB below the Fermi level as depicted in band structures and DOSs calculations. These calculations reveal that graphene features are dominant and the $p$ orbital electrons of carbon are providing the large electron density around the $E_F$ level whereas the $d$-orbital electrons of the Fe have negligible contribution in the total DOSs as shown in the right hand side of Figure 3f. Interestingly, the present DFT computation found that the BLG-Co showed semi-metallic behavior like graphene, although the VB and CB are overlapped onto each other, but the total DOSs calculations show that the electron density is very close to zero at the $E_F$ level (see Figure 3g) like monolayer graphene. The DOSs show that the total DOSs exactly follow the $p$-orbital electron density character of C atoms in graphene and $d$-orbitals electron has negligible contribution indicating that BLG-Co is a semi-metal. The band structures and DOSs calculations of the bulk-BLG-Fe and bulk-BLG-Co are shown in Figure 4f-g. The band structure calculations found that the bands are overlapped and one band crossing point (Dirac point) at the K-point in both the bulk-BLG-Fe and bulk-BLG-Co bulk structure materials. These calculations indicated that the Dirac features were still exist although the TM atoms Fe and Co intercalated in BLG. Electron density appears around the $E_F$ level of both the crystal structure of bulk-BLG-Fe as well as bulk-BLG-Co, and the DOSs calculations showed that $d$-orbitals of the Fe and Co have less contribution in the total DOSs. Thus the graphene features are going to dominate in the crystal structures of the bulk-BLG-Fe and bulk-BLG-Co and the total DOSs follow the same trend of $p$-orbital character of C. Interestingly, our DFT calculations found that VB and CB touch at $E_F$ yielding a Dirac Cone at the H-point in the bulk structure of the BLG-Co material. This result means that the graphene features are dominant around the H-point in the bulk-BLG-Co. Thus we can predict that the bulk-BLG-Co is a Dirac material whereas the BLG-Co is a semi-metal with zero band gap. The DOSs and sub-shells DOSs calculations revealed that *p*-orbital of the carbon in graphene of the BLG-Co, BLG-Fe, bulk-BLG-Co and bulk-BLG-Fe materials are providing the electron density around the Fermi level and the *d*-orbitals of the TM (Co and Fe) has very less contribution in the total DOSs as depicted in Figure 8f-g and 9f-9g, respectively. Thus the graphene features are going to dominate in both the 2D layer structure and bulk structures BLG-Co and BLG-Fe compounds. The sub-shells electron density calculations found that $p_z$ sub-shells of carbon and $d_{yz}$ sub-shells of TM atoms are providing the electron density around the $E_F$ level as displayed in Figure Figure 8f-g and 9f-9g, respectively.

The C-Ni and *d* distances were increased in both the 2D layer and bulk structure BLG when BLG was intercalated by Ni atom in compared to the BLG-Co and bulk-BLG-Co materials. In the case of the BLG-Cu and bulk-BLG-Cu materials, the C-Cu and *d* were changed by 0.02 Å in compared to the BLG-Ni and bulk-BLG-Ni. The electronic properties calculations were carried out in both BLG-Ni, bulk-BLG-Ni, BLG-Cu and bulk-BLG-Cu, including their band structures and DOSs as shown in Figure 3h-i and 4h-i, respectively. The band structures found that the valence and conduction bands overlap on each other and the Dirac point was pushed down below the Fermi level at the K-point of both the BLG-Ni and BLG-Cu. Our present computation reveals that the total DOSs of the BLG-Ni is about zero at the $E_F$, whereas BLG-Cu has large electron density around the $E_F$. The DOSs calculation reveals that BLG-Ni has semi-metallic behavior as there is no electron density at the Fermi level. The total DOSs follow the *p*-orbital character of C atoms of the graphene resulting in the visible dominating graphene features in the BLG-Ni (see the right hand side of Figure 3h). In the case of the BLG-Cu, a large electron density appears in the total DOSs around the Fermi level resulting a conductor due to the overlapping of VB and CB below the $E_F$ at the K-point as depicted in Figure 3i. Electron density, which follows the *p*-orbital character of C atoms of the graphene, occurs around the $E_F$ making it as a Dirac material, and there is no significant contributing components of the *d*-orbital electrons of the Ni as shown in Figure 3h. It can be shown from the sub-shell electrons density analysis. The sub-shell electron density calculations showed that the total density of states are followed by the $p_z$ sub-shells of carbon of the graphene and there is no contributing components of the sub-shells of *d*-orbital of Ni atom as shown in Figure 8h, making BLG-Ni is a semi-metal. A large electron density appears in the total DOSs of the BLG-Cu, and DOSs calculations showed that $p_z$ sub-shells of carbon provides the large electron density as depicted in Figure 8i. The present study revels that the BLG-Cu shows metallic behavior.

The band structures and DOSs of the bulk-BLG-Ni and bulk-BLG-Cu materials are shown in Figure 4h and Figure 4i, respectively. The band structures and DOSs of the bulk-BLG-Ni showed that VB and CB overlap about 1.1 eV above the Fermi level resulting in large electron density around the $E_F$ making it as a conductor shown in Figure 4h. The band calculations showed that the bulk-BLG-Cu the Dirac Cone appeared at the K-point; see Figure 4i. One Dirac point was moved above the $E_F$ at K point and the other one moved down below the $E_F$ of the BLG-Ni crystal, and the total DOSs showed electron density around the $E_F$ due the *p*-orbital electrons of C and the *d*-orbital electrons of Ni. The sub-shell electron density calculations reveals that $p_z$ sub-shells of carbon and $d_{yz}$ sub-shells of Ni are providing the electron density around the $E_F$ level as displayed in Figure 9h. In the BLG-Cu crystal, the Dirac Cone appeared in its band structures at K-point and other Dirac point moved down below the $E_F$ level at H-point as depicted in Figure 9i. Electron density was found around the $E_F$ due the contribution of electron density of $p_z$ sub-shells of carbon as shown in Figure 9i. In both the materials, the VB and CB are overlapped below the $E_F$ level resulting them conductor.

The equilibrium C-Zn bond and intercalation distances of both the BLG-Zn and bulk-BLG-Zn were increased in comparison with all other materials as reported in Table 1. These effects on the electronic properties of both the layer and bulk structure materials. The band structure and DOSs calculations of the 2D layer and bulk structure of BLG-Zn were shown in Figure 3j and Figure 4j, respectively. The Dirac point at K has pushed down below the $E_F$ level and the total DOSs showed large electron distribution around the $E_F$ of the layer structure of the BLG-Zn due to overlap of valence and conduction bands; see Figure 3j. Similarly, the VB and CB of the bulk structure of BLG-Zn overlap above the $E_F$ level resulting in a large electron density around the $E_F$. The present DFT calculation reveals that the total density of states of both the materials (bilayer and bulk structure) follow the *p*-orbital electron density, which also causes the large electron density around the Fermi level making them as a conductor.

# 2 Optimized Structures of BLG-TM materials (.cif format)

The optimized structures are provided below in .cif format.

## 2.1 BLG-Sc: Sc Intercalated 2D Bilayer Graphene

```
data_BLG-Sc_2D_SLAB
_symmetry_space_group_name_H-M    'P6/MMM'
_symmetry_Int_Tables_number       191
_symmetry_cell_setting            hexagonal
loop_
_symmetry_equiv_pos_as_xyz
  x,y,z
  -y,x-y,z
  -x+y,-x,z
  -x,-y,z
  y,-x+y,z
  x-y,x,z
  y,x,-z
  x-y,-y,-z
  -x,-x+y,-z
  -y,-x,-z
  -x+y,y,-z
  x,x-y,-z
  -x,-y,-z
  y,-x+y,-z
  x-y,x,-z
  x,y,-z
  -y,x-y,-z
  -x+y,-x,-z
  -y,-x,z
  -x+y,y,z
  x,x-y,z
  y,x,z
  x-y,-y,z
  -x,-x+y,z
_cell_length_a                    4.9511
_cell_length_b                    4.9511
_cell_length_c                    500.0000
_cell_angle_alpha                 90.0000
_cell_angle_beta                  90.0000
_cell_angle_gamma                 120.0000
loop_
_atom_site_label
_atom_site_type_symbol
_atom_site_fract_x
_atom_site_fract_y
_atom_site_fract_z
_atom_site_U_iso_or_equiv
_atom_site_adp_type
_atom_site_occupancy
C001   C    -0.33614  -0.16807   0.00377   0.00000  Uiso   1.00
C013   C     0.33333  -0.33333   0.00371   0.00000  Uiso   1.00
SC017  Sc    0.00000   0.00000   0.00000   0.00000  Uiso   1.00
```

## 2.2 BLG-Ti: Ti Intercalated 2D Bilayer Graphene

```
data_BLG-Ti_2D_SLAB
_symmetry_space_group_name_H-M    'P6/MMM'
_symmetry_Int_Tables_number       191
_symmetry_cell_setting            hexagonal
loop_
_symmetry_equiv_pos_as_xyz
  x,y,z
  -y,x-y,z
  -x+y,-x,z
  -x,-y,z
  y,-x+y,z
  x-y,x,z
  y,x,-z
  x-y,-y,-z
  -x,-x+y,-z
  -y,-x,-z
  -x+y,y,-z
  x,x-y,-z
  -x,-y,-z
  y,-x+y,-z
  x-y,x,-z
  x,y,-z
  -y,x-y,-z
  -x+y,-x,-z
  -y,-x,z
  -x+y,y,z
  x,x-y,z
  y,x,z
  x-y,-y,z
  -x,-x+y,z
_cell_length_a                    4.9479
_cell_length_b                    4.9479
_cell_length_c                    500.0000
_cell_angle_alpha                 90.0000
_cell_angle_beta                  90.0000
_cell_angle_gamma                 120.0000
loop_
_atom_site_label
_atom_site_type_symbol
_atom_site_fract_x
_atom_site_fract_y
_atom_site_fract_z
_atom_site_U_iso_or_equiv
_atom_site_adp_type
_atom_site_occupancy
C001   C    -0.33662  -0.16831   0.00345   0.00000  Uiso   1.00
C013   C     0.33333  -0.33333   0.00341   0.00000  Uiso   1.00
TI017  Ti    0.00000   0.00000   0.00000   0.00000  Uiso   1.00
```

## 2.3 BLG-V: V Intercalated 2D Bilayer Graphene

```
data_BLG-V_2D_SLAB
_symmetry_space_group_name_H-M    'P6/MMM'
_symmetry_Int_Tables_number       191
_symmetry_cell_setting            hexagonal
loop_
_symmetry_equiv_pos_as_xyz
  x,y,z
  -y,x-y,z
  -x+y,-x,z
  -x,-y,z
  y,-x+y,z
  x-y,x,z
  y,x,-z
  x-y,-y,-z
  -x,-x+y,-z
  -y,-x,-z
  -x+y,y,-z
  x,x-y,-z
  -x,-y,-z
  y,-x+y,-z
  x-y,x,-z
  x,y,-z
  -y,x-y,-z
  -x+y,-x,-z
  -y,-x,z
  -x+y,y,z
  x,x-y,z
  y,x,z
  x-y,-y,z
  -x,-x+y,z
_cell_length_a                    4.9423
_cell_length_b                    4.9423
_cell_length_c                    500.0000
_cell_angle_alpha                 90.0000
_cell_angle_beta                  90.0000
_cell_angle_gamma                 120.0000
loop_
_atom_site_label
_atom_site_type_symbol
_atom_site_fract_x
_atom_site_fract_y
_atom_site_fract_z
_atom_site_U_iso_or_equiv
_atom_site_adp_type
_atom_site_occupancy
C001   C    -0.33601  -0.16801   0.00344   0.00000  Uiso   1.00
C013   C     0.33333  -0.33333   0.00339   0.00000  Uiso   1.00
V017   V     0.00000   0.00000   0.00000   0.00000  Uiso   1.00
```

### 2.4 BLG-Cr: Cr Intercalated 2D Bilayer Graphene

```
data_BLG-Cr_2D_SLAB
_symmetry_space_group_name_H-M    'P6/MMM'
_symmetry_Int_Tables_number       191
_symmetry_cell_setting            hexagonal
loop_
_symmetry_equiv_pos_as_xyz
  x,y,z
  -y,x-y,z
  -x+y,-x,z
  -x,-y,z
  y,-x+y,z
  x-y,x,z
  y,x,-z
  x-y,-y,-z
  -x,-x+y,-z
  -y,-x,-z
  -x+y,y,-z
  x,x-y,-z
  -x,-y,-z
  y,-x+y,-z
  x-y,x,-z
  x,y,-z
  -y,x-y,-z
  -x+y,-x,-z
  -y,-x,z
  -x+y,y,z
  x,x-y,z
  y,x,z
  x-y,-y,z
  -x,-x+y,z
_cell_length_a                    4.9423
_cell_length_b                    4.9423
_cell_length_c                    500.0000
_cell_angle_alpha                 90.0000
_cell_angle_beta                  90.0000
_cell_angle_gamma                 120.0000
loop_
_atom_site_label
_atom_site_type_symbol
_atom_site_fract_x
_atom_site_fract_y
_atom_site_fract_z
_atom_site_U_iso_or_equiv
_atom_site_adp_type
_atom_site_occupancy
C001   C    -0.33587  -0.16794   0.00335   0.00000  Uiso   1.00
C013   C     0.33333  -0.33333   0.00331   0.00000  Uiso   1.00
CR017  Cr    0.00000   0.00000   0.00000   0.00000  Uiso   1.00
```

### 2.5 BLG-Mn: Mn Intercalated 2D Bilayer Graphene

```
data_BLG-Mn_2D_SLAB
_symmetry_space_group_name_H-M    'P6/MMM'
_symmetry_Int_Tables_number       191
_symmetry_cell_setting            hexagonal
loop_
_symmetry_equiv_pos_as_xyz
  x,y,z
  -y,x-y,z
  -x+y,-x,z
  -x,-y,z
  y,-x+y,z
  x-y,x,z
  y,x,-z
  x-y,-y,-z
  -x,-x+y,-z
  -y,-x,-z
  -x+y,y,-z
  x,x-y,-z
  -x,-y,-z
  y,-x+y,-z
  x-y,x,-z
  x,y,-z
  -y,x-y,-z
  -x+y,-x,-z
  -y,-x,z
  -x+y,y,z
  x,x-y,z
  y,x,z
  x-y,-y,z
  -x,-x+y,z
_cell_length_a                    4.9401
_cell_length_b                    4.9401
_cell_length_c                    500.0000
_cell_angle_alpha                 90.0000
_cell_angle_beta                  90.0000
_cell_angle_gamma                 120.0000
loop_
_atom_site_label
_atom_site_type_symbol
_atom_site_fract_x
_atom_site_fract_y
_atom_site_fract_z
_atom_site_U_iso_or_equiv
_atom_site_adp_type
_atom_site_occupancy
C001   C    -0.33577  -0.16789   0.00362   0.00000  Uiso   1.00
C013   C     0.33333  -0.33333   0.00358   0.00000  Uiso   1.00
MN017  Mn    0.00000   0.00000   0.00000   0.00000  Uiso   1.00
```

## 2.6 BLG-Fe: Fe Intercalated 2D Bilayer Graphene

```
data_OPT_BLG-Fe_2D_SLAB
_symmetry_space_group_name_H-M    'P6/MMM'
_symmetry_Int_Tables_number       191
_symmetry_cell_setting            hexagonal
loop_
_symmetry_equiv_pos_as_xyz
  x,y,z
  -y,x-y,z
  -x+y,-x,z
  -x,-y,z
  y,-x+y,z
  x-y,x,z
  y,x,-z
  x-y,-y,-z
  -x,-x+y,-z
  -y,-x,-z
  -x+y,y,-z
  x,x-y,-z
  -x,-y,-z
  y,-x+y,-z
  x-y,x,-z
  x,y,-z
  -y,x-y,-z
  -x+y,-x,-z
  -y,-x,z
  -x+y,y,z
  x,x-y,z
  y,x,z
  x-y,-y,z
  -x,-x+y,z
_cell_length_a                    4.9339
_cell_length_b                    4.9339
_cell_length_c                    500.0000
_cell_angle_alpha                 90.0000
_cell_angle_beta                  90.0000
_cell_angle_gamma                 120.0000
loop_
_atom_site_label
_atom_site_type_symbol
_atom_site_fract_x
_atom_site_fract_y
_atom_site_fract_z
_atom_site_U_iso_or_equiv
_atom_site_adp_type
_atom_site_occupancy
C001   C    -0.33556  -0.16778   0.00349   0.00000  Uiso   1.00
C013   C     0.33333  -0.33333   0.00344   0.00000  Uiso   1.00
FE017  Fe    0.00000   0.00000   0.00000   0.00000  Uiso   1.00
```

## 2.7 BLG-Co: Co Intercalated 2D Bilayer Graphene

```
data_BLG-Co_2D_SLAB
_symmetry_space_group_name_H-M    'P6/MMM'
_symmetry_Int_Tables_number       191
_symmetry_cell_setting            hexagonal
loop_
_symmetry_equiv_pos_as_xyz
  x,y,z
  -y,x-y,z
  -x+y,-x,z
  -x,-y,z
  y,-x+y,z
  x-y,x,z
  y,x,-z
  x-y,-y,-z
  -x,-x+y,-z
  -y,-x,-z
  -x+y,y,-z
  x,x-y,-z
  -x,-y,-z
  y,-x+y,-z
  x-y,x,-z
  x,y,-z
  -y,x-y,-z
  -x+y,-x,-z
  -y,-x,z
  -x+y,y,z
  x,x-y,z
  y,x,z
  x-y,-y,z
  -x,-x+y,z
_cell_length_a                    4.9311
_cell_length_b                    4.9311
_cell_length_c                    500.0000
_cell_angle_alpha                 90.0000
_cell_angle_beta                  90.0000
_cell_angle_gamma                 120.0000
loop_
_atom_site_label
_atom_site_type_symbol
_atom_site_fract_x
_atom_site_fract_y
_atom_site_fract_z
_atom_site_U_iso_or_equiv
_atom_site_adp_type
_atom_site_occupancy
C001   C    -0.33511  -0.16755   0.00345   0.00000  Uiso   1.00
C013   C     0.33333  -0.33333   0.00342   0.00000  Uiso   1.00
CO017  Co    0.00000   0.00000   0.00000   0.00000  Uiso   1.00
```

### 2.8 BLG-Ni: Ni Intercalated 2D Bilayer Graphene

```
data_BLG-Ni_2D_SLAB
_symmetry_space_group_name_H-M    'P6/MMM'
_symmetry_Int_Tables_number       191
_symmetry_cell_setting            hexagonal
loop_
_symmetry_equiv_pos_as_xyz
  x,y,z
  -y,x-y,z
  -x+y,-x,z
  -x,-y,z
  y,-x+y,z
  x-y,x,z
  y,x,-z
  x-y,-y,-z
  -x,-x+y,-z
  -y,-x,-z
  -x+y,y,-z
  x,x-y,-z
  -x,-y,-z
  y,-x+y,-z
  x-y,x,-z
  x,y,-z
  -y,x-y,-z
  -x+y,-x,-z
  -y,-x,z
  -x+y,y,z
  x,x-y,z
  y,x,z
  x-y,-y,z
  -x,-x+y,z
_cell_length_a                    4.9244
_cell_length_b                    4.9244
_cell_length_c                    500.0000
_cell_angle_alpha                 90.0000
_cell_angle_beta                  90.0000
_cell_angle_gamma                 120.0000
loop_
_atom_site_label
_atom_site_type_symbol
_atom_site_fract_x
_atom_site_fract_y
_atom_site_fract_z
_atom_site_U_iso_or_equiv
_atom_site_adp_type
_atom_site_occupancy
C001   C    -0.33492  -0.16746   0.00351   0.00000  Uiso   1.00
C013   C     0.33333  -0.33333   0.00350   0.00000  Uiso   1.00
NI017  Ni    0.00000   0.00000   0.00000   0.00000  Uiso   1.00
```

## 2.9 BLG-Cu: Cu Intercalated 2D Bilayer Graphene

```
data_BLG-Cu_2D_SLAB
_symmetry_space_group_name_H-M    'P6/MMM'
_symmetry_Int_Tables_number       191
_symmetry_cell_setting            hexagonal
loop_
_symmetry_equiv_pos_as_xyz
  x,y,z
  -y,x-y,z
  -x+y,-x,z
  -x,-y,z
  y,-x+y,z
  x-y,x,z
  y,x,-z
  x-y,-y,-z
  -x,-x+y,-z
  -y,-x,-z
  -x+y,y,-z
  x,x-y,-z
  -x,-y,-z
  y,-x+y,-z
  x-y,x,-z
  x,y,-z
  -y,x-y,-z
  -x+y,-x,-z
  -y,-x,z
  -x+y,y,z
  x,x-y,z
  y,x,z
  x-y,-y,z
  -x,-x+y,z
_cell_length_a                    4.9270
_cell_length_b                    4.9270
_cell_length_c                    500.0000
_cell_angle_alpha                 90.0000
_cell_angle_beta                  90.0000
_cell_angle_gamma                 120.0000
loop_
_atom_site_label
_atom_site_type_symbol
_atom_site_fract_x
_atom_site_fract_y
_atom_site_fract_z
_atom_site_U_iso_or_equiv
_atom_site_adp_type
_atom_site_occupancy
C001   C    -0.33500  -0.16750   0.00353   0.00000  Uiso   1.00
C013   C     0.33333  -0.33333   0.00349   0.00000  Uiso   1.00
CU017  Cu    0.00000   0.00000   0.00000   0.00000  Uiso   1.00
```

## 2.10 BLG-Zn: Zn Intercalated 2D Bilayer Graphene

```
data_BLG-Zn_2D_SLAB
_symmetry_space_group_name_H-M    'P6/MMM'
_symmetry_Int_Tables_number       191
_symmetry_cell_setting            hexagonal
loop_
_symmetry_equiv_pos_as_xyz
  x,y,z
  -y,x-y,z
  -x+y,-x,z
  -x,-y,z
  y,-x+y,z
  x-y,x,z
  y,x,-z
  x-y,-y,-z
  -x,-x+y,-z
  -y,-x,-z
  -x+y,y,-z
  x,x-y,-z
  -x,-y,-z
  y,-x+y,-z
  x-y,x,-z
  x,y,-z
  -y,x-y,-z
  -x+y,-x,-z
  -y,-x,z
  -x+y,y,z
  x,x-y,z
  y,x,z
  x-y,-y,z
  -x,-x+y,z
_cell_length_a                    4.8988
_cell_length_b                    4.8988
_cell_length_c                    500.0000
_cell_angle_alpha                 90.0000
_cell_angle_beta                  90.0000
_cell_angle_gamma                 120.0000
loop_
_atom_site_label
_atom_site_type_symbol
_atom_site_fract_x
_atom_site_fract_y
_atom_site_fract_z
_atom_site_U_iso_or_equiv
_atom_site_adp_type
_atom_site_occupancy
C001   C    -0.33346  -0.16673   0.00578   0.00000  Uiso   1.00
C013   C     0.33333  -0.33333   0.00578   0.00000  Uiso   1.00
ZN017  Zn    0.00000   0.00000   0.00000   0.00000  Uiso   1.00
```

# 3 Optimized Structures of Bulk-BLG-TM Materials (.cif format)

The optimized structures are provided below in .cif format.

## 3.1 Bulk-BLG-Sc: Sc Intercalated Bulk Structure BLG

```
data_BLG-Sc_3D_Bulk_Graphite
_symmetry_space_group_name_H-M    'P6/MMM'
_symmetry_Int_Tables_number       191
_symmetry_cell_setting            hexagonal
loop_
_symmetry_equiv_pos_as_xyz
  x,y,z
  -y,x-y,z
  -x+y,-x,z
  -x,-y,z
  y,-x+y,z
  x-y,x,z
  y,x,-z
  x-y,-y,-z
  -x,-x+y,-z
  -y,-x,-z
  -x+y,y,-z
  x,x-y,-z
  -x,-y,-z
  y,-x+y,-z
  x-y,x,-z
  x,y,-z
  -y,x-y,-z
  -x+y,-x,-z
  -y,-x,z
  -x+y,y,z
  x,x-y,z
  y,x,z
  x-y,-y,z
  -x,-x+y,z
_cell_length_a                    4.9861
_cell_length_b                    4.9861
_cell_length_c                    3.8001
_cell_angle_alpha                 90.0000
_cell_angle_beta                  90.0000
_cell_angle_gamma                 120.0000
loop_
_atom_site_label
_atom_site_type_symbol
_atom_site_fract_x
_atom_site_fract_y
_atom_site_fract_z
_atom_site_U_iso_or_equiv
_atom_site_adp_type
_atom_site_occupancy
C001   C     0.16850   0.33700   0.00000   0.00000  Uiso   1.00
C007   C     0.33333  -0.33333   0.00000   0.00000  Uiso   1.00
SC009  Sc   -0.00000   0.00000  -0.50000   0.00000  Uiso   1.00
```

### 3.2 Bulk-BLG-Ti: Ti Intercalated Bulk Structure BLG

```
data_BLG-Ti_3D_Bulk_Graphite
_symmetry_space_group_name_H-M    'P6/MMM'
_symmetry_Int_Tables_number       191
_symmetry_cell_setting            hexagonal
loop_
_symmetry_equiv_pos_as_xyz
  x,y,z
  -y,x-y,z
  -x+y,-x,z
  -x,-y,z
  y,-x+y,z
  x-y,x,z
  y,x,-z
  x-y,-y,-z
  -x,-x+y,-z
  -y,-x,-z
  -x+y,y,-z
  x,x-y,-z
  -x,-y,-z
  y,-x+y,-z
  x-y,x,-z
  x,y,-z
  -y,x-y,-z
  -x+y,-x,-z
  -y,-x,z
  -x+y,y,z
  x,x-y,z
  y,x,z
  x-y,-y,z
  -x,-x+y,z
_cell_length_a                    4.9956
_cell_length_b                    4.9956
_cell_length_c                    3.4253
_cell_angle_alpha                 90.0000
_cell_angle_beta                  90.0000
_cell_angle_gamma                 120.0000
loop_
_atom_site_label
_atom_site_type_symbol
_atom_site_fract_x
_atom_site_fract_y
_atom_site_fract_z
_atom_site_U_iso_or_equiv
_atom_site_adp_type
_atom_site_occupancy
C001   C     0.16983   0.33967  -0.00000   0.00000  Uiso   1.00
C007   C     0.33333  -0.33333   0.00000   0.00000  Uiso   1.00
TI009  Ti   -0.00000   0.00000   0.50000   0.00000  Uiso   1.00
```

### 3.3 Bulk-BLG-V: V Intercalated Bulk Structure BLG

```
data_BLG-V_3D_Bulk_Graphite
_symmetry_space_group_name_H-M    'P6/MMM'
_symmetry_Int_Tables_number       191
_symmetry_cell_setting            hexagonal
loop_
_symmetry_equiv_pos_as_xyz
  x,y,z
  -y,x-y,z
  -x+y,-x,z
  -x,-y,z
  y,-x+y,z
  x-y,x,z
  y,x,-z
  x-y,-y,-z
  -x,-x+y,-z
  -y,-x,-z
  -x+y,y,-z
  x,x-y,-z
  -x,-y,-z
  y,-x+y,-z
  x-y,x,-z
  x,y,-z
  -y,x-y,-z
  -x+y,-x,-z
  -y,-x,z
  -x+y,y,z
  x,x-y,z
  y,x,z
  x-y,-y,z
  -x,-x+y,z
_cell_length_a                    4.9909
_cell_length_b                    4.9909
_cell_length_c                    3.4500
_cell_angle_alpha                 90.0000
_cell_angle_beta                  90.0000
_cell_angle_gamma                 120.0000
loop_
_atom_site_label
_atom_site_type_symbol
_atom_site_fract_x
_atom_site_fract_y
_atom_site_fract_z
_atom_site_U_iso_or_equiv
_atom_site_adp_type
_atom_site_occupancy
C001   C     0.16872   0.33745  -0.00000   0.00000  Uiso   1.00
C007   C     0.33333  -0.33333  -0.00000   0.00000  Uiso   1.00
V009   V    -0.00000   0.00000   0.50000   0.00000  Uiso   1.00
```

### 3.4 Bulk-BLG-Cr: Cr Intercalated Bulk Structure BLG

```
data_BLG-Cr_3D_Bulk_Graphite
_symmetry_space_group_name_H-M    'P6/MMM'
_symmetry_Int_Tables_number       191
_symmetry_cell_setting            hexagonal
loop_
_symmetry_equiv_pos_as_xyz
  x,y,z
  -y,x-y,z
  -x+y,-x,z
  -x,-y,z
  y,-x+y,z
  x-y,x,z
  y,x,-z
  x-y,-y,-z
  -x,-x+y,-z
  -y,-x,-z
  -x+y,y,-z
  x,x-y,-z
  -x,-y,-z
  y,-x+y,-z
  x-y,x,-z
  x,y,-z
  -y,x-y,-z
  -x+y,-x,-z
  -y,-x,z
  -x+y,y,z
  x,x-y,z
  y,x,z
  x-y,-y,z
  -x,-x+y,z
_cell_length_a                    4.9800
_cell_length_b                    4.9800
_cell_length_c                    3.4708
_cell_angle_alpha                 90.0000
_cell_angle_beta                  90.0000
_cell_angle_gamma                 120.0000
loop_
_atom_site_label
_atom_site_type_symbol
_atom_site_fract_x
_atom_site_fract_y
_atom_site_fract_z
_atom_site_U_iso_or_equiv
_atom_site_adp_type
_atom_site_occupancy
C001   C     0.16904   0.33809  -0.00000   0.00000  Uiso   1.00
C007   C     0.33333  -0.33333   0.00000   0.00000  Uiso   1.00
CR009  Cr   -0.00000   0.00000  -0.50000   0.00000  Uiso   1.00
```

### 3.5 Bulk-BLG-Mn: Mn Intercalated Bulk Structure BLG

```
data_BLG-Mn_3D_Bulk_Graphite
_symmetry_space_group_name_H-M    'P6/MMM'
_symmetry_Int_Tables_number       191
_symmetry_cell_setting            hexagonal
loop_
_symmetry_equiv_pos_as_xyz
  x,y,z
  -y,x-y,z
  -x+y,-x,z
  -x,-y,z
  y,-x+y,z
  x-y,x,z
  y,x,-z
  x-y,-y,-z
  -x,-x+y,-z
  -y,-x,-z
  -x+y,y,-z
  x,x-y,-z
  -x,-y,-z
  y,-x+y,-z
  x-y,x,-z
  x,y,-z
  -y,x-y,-z
  -x+y,-x,-z
  -y,-x,z
  -x+y,y,z
  x,x-y,z
  y,x,z
  x-y,-y,z
  -x,-x+y,z
_cell_length_a                    4.9799
_cell_length_b                    4.9799
_cell_length_c                    3.5837
_cell_angle_alpha                 90.0000
_cell_angle_beta                  90.0000
_cell_angle_gamma                 120.0000
loop_
_atom_site_label
_atom_site_type_symbol
_atom_site_fract_x
_atom_site_fract_y
_atom_site_fract_z
_atom_site_U_iso_or_equiv
_atom_site_adp_type
_atom_site_occupancy
C001   C     0.16884   0.33769   0.00000   0.00000  Uiso   1.00
C007   C     0.33333  -0.33333  -0.00000   0.00000  Uiso   1.00
MN009  Mn    0.00000   0.00000  -0.50000   0.00000  Uiso   1.00
```

### 3.6 Bulk-BLG-Fe: Fe Intercalated Bulk Structure BLG

```
data_BLG-Fe_3D_Bulk_Graphite
_symmetry_space_group_name_H-M    'P6/MMM'
_symmetry_Int_Tables_number       191
_symmetry_cell_setting            hexagonal
loop_
_symmetry_equiv_pos_as_xyz
  x,y,z
  -y,x-y,z
  -x+y,-x,z
  -x,-y,z
  y,-x+y,z
  x-y,x,z
  y,x,-z
  x-y,-y,-z
  -x,-x+y,-z
  -y,-x,-z
  -x+y,y,-z
  x,x-y,-z
  -x,-y,-z
  y,-x+y,-z
  x-y,x,-z
  x,y,-z
  -y,x-y,-z
  -x+y,-x,-z
  -y,-x,z
  -x+y,y,z
  x,x-y,z
  y,x,z
  x-y,-y,z
  -x,-x+y,z
_cell_length_a                    4.9733
_cell_length_b                    4.9733
_cell_length_c                    3.4230
_cell_angle_alpha                 90.0000
_cell_angle_beta                  90.0000
_cell_angle_gamma                 120.0000
loop_
_atom_site_label
_atom_site_type_symbol
_atom_site_fract_x
_atom_site_fract_y
_atom_site_fract_z
_atom_site_U_iso_or_equiv
_atom_site_adp_type
_atom_site_occupancy
C001   C     0.16899   0.33797  -0.00000   0.00000  Uiso   1.00
C007   C     0.33333  -0.33333  -0.00000   0.00000  Uiso   1.00
FE009  Fe    0.00000   0.00000  -0.50000   0.00000  Uiso   1.00
```

### 3.7 Bulk-BLG-Co: Co Intercalated Bulk Structure BLG

```
data_BLG-Co_3D_Bulk_Graphite
_symmetry_space_group_name_H-M    'P6/MMM'
_symmetry_Int_Tables_number       191
_symmetry_cell_setting            hexagonal
loop_
_symmetry_equiv_pos_as_xyz
  x,y,z
  -y,x-y,z
  -x+y,-x,z
  -x,-y,z
  y,-x+y,z
  x-y,x,z
  y,x,-z
  x-y,-y,-z
  -x,-x+y,-z
  -y,-x,-z
  -x+y,y,-z
  x,x-y,-z
  -x,-y,-z
  y,-x+y,-z
  x-y,x,-z
  x,y,-z
  -y,x-y,-z
  -x+y,-x,-z
  -y,-x,z
  -x+y,y,z
  x,x-y,z
  y,x,z
  x-y,-y,z
  -x,-x+y,z
_cell_length_a                    4.9657
_cell_length_b                    4.9657
_cell_length_c                    3.3684
_cell_angle_alpha                 90.0000
_cell_angle_beta                  90.0000
_cell_angle_gamma                 120.0000
loop_
_atom_site_label
_atom_site_type_symbol
_atom_site_fract_x
_atom_site_fract_y
_atom_site_fract_z
_atom_site_U_iso_or_equiv
_atom_site_adp_type
_atom_site_occupancy
C001   C     0.16820   0.33639  -0.00000   0.00000  Uiso   1.00
C007   C     0.33333  -0.33333   0.00000   0.00000  Uiso   1.00
CO009  Co   -0.00000   0.00000   0.50000   0.00000  Uiso   1.00
```

### 3.8 Bulk-BLG-Ni: Ni Intercalated Bulk Structure BLG

```
data_BLG-Ni_3D_Bulk_Graphite
_symmetry_space_group_name_H-M    'P6/MMM'
_symmetry_Int_Tables_number       191
_symmetry_cell_setting            hexagonal
loop_
_symmetry_equiv_pos_as_xyz
  x,y,z
  -y,x-y,z
  -x+y,-x,z
  -x,-y,z
  y,-x+y,z
  x-y,x,z
  y,x,-z
  x-y,-y,-z
  -x,-x+y,-z
  -y,-x,-z
  -x+y,y,-z
  x,x-y,-z
  -x,-y,-z
  y,-x+y,-z
  x-y,x,-z
  x,y,-z
  -y,x-y,-z
  -x+y,-x,-z
  -y,-x,z
  -x+y,y,z
  x,x-y,z
  y,x,z
  x-y,-y,z
  -x,-x+y,z
_cell_length_a                    4.9531
_cell_length_b                    4.9531
_cell_length_c                    3.4085
_cell_angle_alpha                 90.0000
_cell_angle_beta                  90.0000
_cell_angle_gamma                 120.0000
loop_
_atom_site_label
_atom_site_type_symbol
_atom_site_fract_x
_atom_site_fract_y
_atom_site_fract_z
_atom_site_U_iso_or_equiv
_atom_site_adp_type
_atom_site_occupancy
C001   C     0.16825   0.33650   0.00000   0.00000  Uiso   1.00
C007   C     0.33333  -0.33333  -0.00000   0.00000  Uiso   1.00
NI009  Ni   -0.00000  -0.00000  -0.50000   0.00000  Uiso   1.00
```

### 3.9 Bulk-BLG-Cu: Cu Intercalated Bulk Structure BLG

```
data_BLG-Co_3D_Bulk_Graphite
_symmetry_space_group_name_H-M    'P6/MMM'
_symmetry_Int_Tables_number       191
_symmetry_cell_setting            hexagonal
loop_
_symmetry_equiv_pos_as_xyz
  x,y,z
  -y,x-y,z
  -x+y,-x,z
  -x,-y,z
  y,-x+y,z
  x-y,x,z
  y,x,-z
  x-y,-y,-z
  -x,-x+y,-z
  -y,-x,-z
  -x+y,y,-z
  x,x-y,-z
  -x,-y,-z
  y,-x+y,-z
  x-y,x,-z
  x,y,-z
  -y,x-y,-z
  -x+y,-x,-z
  -y,-x,z
  -x+y,y,z
  x,x-y,z
  y,x,z
  x-y,-y,z
  -x,-x+y,z
_cell_length_a                    4.9595
_cell_length_b                    4.9595
_cell_length_c                    3.4128
_cell_angle_alpha                 90.0000
_cell_angle_beta                  90.0000
_cell_angle_gamma                 120.0000
loop_
_atom_site_label
_atom_site_type_symbol
_atom_site_fract_x
_atom_site_fract_y
_atom_site_fract_z
_atom_site_U_iso_or_equiv
_atom_site_adp_type
_atom_site_occupancy
C001   C     0.16853   0.33705   0.00000   0.00000  Uiso   1.00
C007   C     0.33333  -0.33333   0.00000   0.00000  Uiso   1.00
CU009  Cu   -0.00000  -0.00000   0.50000   0.00000  Uiso   1.00
```

### 3.10 Bulk-BLG-Zn: Zn Intercalated Bulk Structure BLG

```
data_BLG-Zn_3D_Bulk_Graphite
_symmetry_space_group_name_H-M    'P6/MMM'
_symmetry_Int_Tables_number       191
_symmetry_cell_setting            hexagonal
loop_
_symmetry_equiv_pos_as_xyz
  x,y,z
  -y,x-y,z
  -x+y,-x,z
  -x,-y,z
  y,-x+y,z
  x-y,x,z
  y,x,-z
  x-y,-y,-z
  -x,-x+y,-z
  -y,-x,-z
  -x+y,y,-z
  x,x-y,-z
  -x,-y,-z
  y,-x+y,-z
  x-y,x,-z
  x,y,-z
  -y,x-y,-z
  -x+y,-x,-z
  -y,-x,z
  -x+y,y,z
  x,x-y,z
  y,x,z
  x-y,-y,z
  -x,-x+y,z
_cell_length_a                    4.8944
_cell_length_b                    4.8944
_cell_length_c                    5.7013
_cell_angle_alpha                 90.0000
_cell_angle_beta                  90.0000
_cell_angle_gamma                 120.0000
loop_
_atom_site_label
_atom_site_type_symbol
_atom_site_fract_x
_atom_site_fract_y
_atom_site_fract_z
_atom_site_U_iso_or_equiv
_atom_site_adp_type
_atom_site_occupancy
C001   C     0.16682   0.33364   0.00000   0.00000  Uiso   1.00
C007   C     0.33333  -0.33333   0.00000   0.00000  Uiso   1.00
ZN009  Zn    0.00000   0.00000  -0.50000   0.00000  Uiso   1.00
```